\documentclass[fleqn,usenatbib]{mnras}

\usepackage{newtxtext,newtxmath}

\usepackage[T1]{fontenc}
\usepackage{ae,aecompl}
\usepackage{natbib}

\usepackage{graphicx}	
\usepackage{amsmath}	
\usepackage{amssymb}	

\title[GRAWITA follow-up of GW150914 and GW151226]
{GRAWITA: VLT Survey Telescope observations of the\\ 
 gravitational wave sources GW150914 and GW151226 }

\author[E. Brocato et al.]
{E. Brocato$^{1}$\thanks{\vspace {-1.5cm} E-mail: enzo.brocato@oa-roma.inaf.it},
M. Branchesi$^{2,3}$,
E. Cappellaro$^{4}$, 
S. Covino$^{5}$,
A. Grado$^{6}$,
G. Greco$^{2,3}$,
\newauthor
L. Limatola$^{6}$,
G. Stratta$^{2,3}$,
S. Yang$^{4}$,
S. Campana$^{5}$,
P. D'Avanzo$^{5}$,
F. Getman$^{6}$,
\newauthor
A. Melandri$^{5}$,
L. Nicastro$^{7}$,
E. Palazzi$^{7}$,
E. Pian$^{7,8}$,
S. Piranomonte$^{1}$,
L. Pulone$^{1}$,
\newauthor
A. Rossi$^{7}$,
L. Tomasella$^{,4}$,
L. Amati,$^{7}$,
L. A. Antonelli$^{1}$,
S. Ascenzi $^{1}$
S. Benetti$^{4}$,
\newauthor
A. Bulgarelli$^{7}$,
M. Capaccioli$^{9}$,
G. Cella$^{8}$,
M. Dadina$^{7}$,
G. De Cesare$^{7}$,
V. D'Elia$^{1}$, 
\newauthor
G. Ghirlanda$^{5}$,
G. Ghisellini$^{5}$,
G. Giuffrida$^{1}$,
G. Iannicola$^{1}$,
G. Israel$^{1}$,
M. Lisi$^{1}$,
\newauthor
F. Longo$^{10}$,
M. Mapelli$^{,4}$,
S. Marinoni$^{1}$,
P. Marrese$^{1}$,
N. Masetti$^{7,}$$^{11}$,
B. Patricelli$^{8}$,
\newauthor
A. Possenti$^{12}$,
M. Radovich$^{4}$,
M. Razzano$^{8}$,
R. Salvaterra$^{13}$,
P. Schipani$^{6}$,
M. Spera$^{4}$,
A. Stamerra$^{8}$, 
\newauthor
L. Stella$^{1}$,
G. Tagliaferri$^{5}$,
V. Testa$^{1}$
\\
\hspace {5.5cm}{\sl GRAWITA - GRAvitational Wave Inaf TeAm}
\\
$^{1}$ INAF, Osservatorio Astronomico di Roma, Via di Frascati, 33, I-00078 Monteporzio Catone, Italy\\
$^{2}$ Universit\`a degli Studi di Urbino `Carlo Bo', Dipartimento di Scienze Pure e Applicate, P.za Repubblica 13, I-61029, Urbino, Italy\\
$^{3}$ INFN, Sezione di Firenze, I-50019 Sesto Fiorentino, Firenze, Italy\\
$^{4}$ INAF, Osservatorio Astronomico di Padova, Vicolo dell'Osservatorio 5, I-35122 Padova, Italy\\
$^{5}$ INAF, Osservatorio Astronomico di Brera, Via E. Bianchi 46, I-23807 Merate (LC), Italy\\
$^{6}$ INAF, Osservatorio Astronomico di Capodimonte, salita Moiariello 16, I-80131, Napoli, Italy\\
$^{7}$ INAF, Istituto di Astrofisica Spaziale e Fisica Cosmica di Bologna, Via Gobetti 101, I-40129 Bologna, Italy\\
$^{8}$ Scuola Normale Superiore, Piazza dei Cavalieri 7, I-56126 Pisa, Italy\\
$^{9}$ Universit\`a degli Studi di Napoli  `Federico II', Dipartimento di Fisica, 
Via Cinthia 21, I-80126 - Napoli, Italy\\
$^{10}$ Universit\`a degli Studi di Trieste e INFN, via Valerio 2, 34127 Trieste, Italy\\
$^{11}$ Departamento de Ciencias F\"isicas, Universidad Andr\"es Bello, Fern\"andez Concha 700, Las Condes, Santiago, Chile\\
$^{12}$ INAF, Osservatorio Astronomico di Cagliari, Via della Scienza 5, I-09047 Selargius (CA), Italy\\
$^{13}$ INAF, Istituto di Astrofisica Spaziale e Fisica Cosmica di Milano, via E. Bassini 15, I-20133 Milano, Italy
} 
\date{Accepted ... Received ...; in original form ...}
\pubyear{2017}
\begin{document}
\label{firstpage}
\pagerange{\pageref{firstpage}--\pageref{lastpage}}
\maketitle
\begin{abstract}
We report the results of deep optical follow-up surveys of the first two gravitational-wave sources, GW150914 and GW151226, done by the GRAvitational Wave Inaf TeAm Collaboration (GRAWITA). The VLT Survey Telescope (VST) responded promptly to the gravitational-wave alerts sent by the LIGO and Virgo Collaborations, monitoring a region of  $90$ deg$^2$ and $72$ deg$^2$ for GW150914 and GW151226, respectively, and repeated the observations over nearly two months. Both surveys reached an average limiting magnitude of about 21 in the $r-$band. The paper describes the VST observational strategy and two independent procedures developed to search for transient counterpart candidates in multi-epoch VST images. Several transients have been discovered but no candidates are recognized to be related to the gravitational-wave events. Interestingly, among many contaminant supernovae, we find a possible correlation between the supernova VSTJ57.77559-59.13990 and GRB\,150827A detected by {\it Fermi}-GBM. The detection efficiency of VST observations for different types of electromagnetic counterparts of gravitational-wave events are evaluated for the present and future follow-up surveys. 
\end{abstract}




\section{Introduction}

The existence of gravitational waves (GWs) has been predicted by the theory of general relativity one century ago as perturbations of space time metric produced by rapidly accelerating quadrupole mass distribution \citep{Einstein1, Einstein2}. GWs are emitted with detectable amplitude by different kinds of astrophysical sources. Among those, coalescence of binary systems of compact objects such as two neutron stars (BNS), a NS and a stellar-mass black hole (NSBH) or two black holes (BBH), collapse of massive stars with large degree of asymmetry and fast rotating asymmetric isolated NSs are expected to emit in the sensitive frequency range (10 Hz - 10 kHz) of the present generation of GW detectors.

{In September 2015, the longstanding search for gravitational radiation was finally accomplished with the detection by the LIGO and Virgo Collaboration (LVC) of unambiguous emission of GW radiation from an astrophysical source. After detailed analysis, it was recognized that the emission was originated in the coalescence of two BHs at a cosmological redshift of $z\simeq{}0.09$ \citep{Abbott2016a}. Two months later, at the end of December 2015,  {the GWs emitted by a second BBH system, again at $z\simeq{}0.09$, were detected \citep{Abbott2016g}. The discoveries were carried out by the two US-based Advanced LIGO observatories  \citep[aLIGO,][]{Aasi2015}, a network of two 4-km length laser interferometers located in Hanford (Washington) and Livingston (Louisiana), respectively. The sky localization of GW signals with 2-site network, like the aLIGO, spans from a few hundreds to thousand of square degrees \citep[][]{Singer2014,Essick2015}. The large sky region to observe is the major challenge for the search and identification of possibly associated electromagnetic (EM) emission.

Based on our current understanding, stellar-mass BBH are not expected to produce detectable EM emission due to the absence of accreting 
material\footnote{Only recently some mechanisms that could produce unusual presence of matter around BHs have been discussed \citep{Loeb2016,Perna2016,Zhang2016,bartos17,deMink17}
suggesting that the merger of a BBH is associated with an EM counterpart under particular circumstances.}.
However, if a counterpart} is found, a wealth of important information can be obtained. For example, source localization to arcmin/arcsec level, depending on the observation wavelength, may enable to localize the possible host galaxy.
Spectroscopic redshift can provide an independent estimate of the distance of the source as well as a characterization of the interstellar environment where the source is embedded (e.g. chemical enrichment and ionization status, etc.), thus providing additional information on the source nature and evolutionary history. Part of this information can be used as priors in the GW data analysis and parameter estimation processes.   

The potential gain of detecting the EM counterpart of GW transients motivated a world-wide effort of the whole astronomical community, employing many telescopes and instruments, ground and space-based, ranging from high energy through optical to radio wavelengths, each contributing the monitoring of a portion of the sky localization area with different depth and cadence\footnote{See program description at $~~~~~~~~~~~~~~~~$ http://www.ligo.org/scientists/GWEMalerts.php.}.

In this paper we describe the observational campaign performed by the GRAvitational Wave INAF TeAm (GRAWITA) to follow up the GW triggers during the first LVC  scientific run (O1)  by using the ESO-VLT Survey Telescope (VST), its results and the prospects for the upcoming years. 
In section~\ref{obs} some details on the VST telescope  and the observational strategy  are presented, including  the specific observational response to the  LVC triggers  GW150914 and GW151226. A brief summary of the adopted pre-reduction is described in section~\ref{dataproc}. In the same section, we present  our  approach to the  transient search and, in particular the two independent pipelines (ph-pipe and the diff-pipe) we developed to this aim.  In the following section~\ref{result-foll}, the results of the search are described. For each of the two GW alerts, a subsection is first dedicated to the  previously discovered SNe then the list of transient candidates is discussed.
In section~\ref{upplimits} we describe the upper limits for different types of GW counterpart, which can be obtained from our VST observations. A brief discussion will close the paper (section~\ref{concl}).

\section{VST observational strategy}
\label{obs}

\begin{table*}
\parbox{.45\linewidth}{
\centering
\caption{\label{data} Epochs and dates of the VST observations performed for the GW150914 event. The covered area and the night average seeing full width half maximum are reported in the last two columns.}
\begin{tabular}{lcccc}
\\
\hline\hline
 &    ~~~~~~~  GW150914 &
\\
\hline\hline
Epoch & Date & Area & FWHM\\
     &   (UT)   &    deg$^2$    &  arcsec \\
\hline
1       & 2015-09-17 & 54 & 0.9\\ 
2       & 2015-09-18 & 90 & 0.9\\ 
3       & 2015-09-21 & 90 & 0.9\\ 
4       & 2015-09-25 & 90 & 1.1\\ 
5       & 2015-10-01 & 72 & 1.0\\ 
5       & 2015-10-03 & 18 & 1.0\\ 
6       & 2015-10-14 & 45 & 1.5\\ 
6       & 2015-11-16 &  9 & 1.2\\ 
6       & 2015-11-17 & 18 & 1.1\\ 
6       & 2015-11-18 & 18 & 1.5\\ 
\hline
\end{tabular}
}
\hfill
\parbox{.45\linewidth}{
\centering
\caption{\label{data2} Epochs and dates of the VST observations performed for the GW151226 event. The covered area and the night average seeing full width half maximum are reported in the last two columns.}
\begin{tabular}{lcccc}
\\
\hline\hline
 &    ~~~~~~~  GW151226 &
\\
\hline\hline
Epoch & Date & Area & FWHM\\
     &   (UT)   &    deg$^2$    &  arcsec \\
\hline
1       & 2015-12-27 & 72 & 1.0\\ 
2       & 2015-12-29 & 72 & 1.6\\ 
3       & 2015-12-30 & 9 & 1.3\\ 
3       & 2016-01-01 & 45 & 0.9\\ 
3       & 2016-01-02 & 9 & 0.9\\ 
4       & 2016-01-05 & 18 & 1.2\\ 
4       & 2016-01-06 & 18 & 1.1\\ 
4       & 2016-01-07 & 27 & 0.8\\ 
5       & 2016-01-13 & 45 & 1.5\\ 
5       & 2016-01-14 & 27 & 1.1\\ 
6       & from 2016-01-28 &  & \\
        & to 2016-02-10 & 63 & 1.1\\
\hline
\end{tabular}
}
\end{table*}

The LVC carried out the first observing run (O1) from September 2015 to January 2016, providing three alerts for GW transient candidates (one subsequently determined not to be a viable GW candidate) that were reported to the team of observers participating in the LVC EM follow-up program. 

The first GW candidate alert was sent on 16 September 2015. After the real-time processing of data from LIGO Hanford Observatory (H1) and LIGO Livingston Observatory (L1), an event occurred on 14 September 2015 at 09:50:45 UTC was identified  
\citep[][]{GCN18330}. 

GW150914 was immediately considered an event of interest because the false alarm rate (FAR) threshold determined by the online analysis  passed the alert threshold of 1 per month adopted for O1. 
Further analysis showed that the GW event was produced by the coalescence of two black holes with rest frame masses of $29^{+4}_{-4}$M$_{\odot}$ and $36^{+5}_{-4} $M$_{\odot}$ at a luminosity distance of  $410^{+160}_{-180}$ Mpc \citep{Abbott2016e}. This information became available only months after the trigger, that is, after completion of the EM follow up campaign.
Twenty-five teams of astronomers promptly reacted to the alert and 
an extensive electromagnetic follow-up campaign and archival searches were performed covering the whole electromagnetic spectrum \citep{Abbott2016f, Abbott2016z}.

On 26 December 2015, a further GW candidate (GW151226) was observed by LVC  \citep[][]{GCN18728}.
Again, the GW event resulted from the coalescence of two black holes of rest frame masses of $14.2^{+8.3}_{-3.7}$ M$_{\odot}$ and $7.5\pm2.3$ M$_{\odot}$
at a distance of $440^{+180}_{-190}$ Mpc \citep{Abbott2016g}. The multi-messenger follow-up started on 27 December 2015, more than 1 day after the GW trigger \citep[][]{GCN18728}, again with an excellent response from astronomers' community.

For the search of possible associated optical transients, our team exploited the ESO VST, a 2.6m, 1 deg$^2$ field of view (FoV) imaging telescope located at the Cerro Paranal Observatory in Chile \citep{Capaccioli2011,Kuijken2011} and dedicated to large sky surveys in the austral hemisphere.  The telescope optical design allows to achieve a uniform PSF with variation $<4\%$ over the whole field of view. 
The VST is equipped with the OmegaCAM camera, which covers the field of view of 1 square degree with a scale of 0.21 arcsec/pixel, through a mosaic of 32 CCDs.

The required time allocation was obtained in the framework of the Guarantee Time Observations (GTO) assigned by ESO to the telescope and camera teams in reward of their effort for the construction of the instrument. 
The planned strategy of the follow up transient survey foresees to monitor a sky area of up to 100 deg$^2$ at 5/6 different epochs beginning soon after the GW trigger and lasting 8-10 weeks.

With the announcement of each trigger, different 
low-latency probability sky maps\footnote{FITS format files containing HEALPix (Hierarchical Equal Area isoLatitude Pixelization) sky projection, where to each pixel is assigned the probability to find the GW source in that position of the sky.} were distributed to the teams of observers \citep[][]{GCN18728,GCN18330}. For GW150914  two initial sky maps were produced by un-modelled  searches for GW bursts, one by the coherent Wave Burst (cWB) pipeline \citep{Klimenko2016} and the other by the Bayesian inference algorithm LALInferenceBurst (LIB) \citep{Essick2015}. The cWB and LIB sky maps encompass a 90\% confidence region of 310 deg$^2$ and 750 deg$^2$, respectively. For GW151226, the initial localization was generated by the Bayesian  localization algorithm BAYESTAR \citep{singer-price}. The BAYESTAR sky map  encompasses a 90\% confidence region of 1400 deg$^2$.

In O1, the LVC alerts were not accompanied by information on the source properties such as distance and source type. 
We choose the cWB skymap for GW150914 and the BAYESTAR skymap for GW151226, and planned our observing strategy to maximize the contained probability of GW localization accessible during the Paranal night.
For the temporal sampling, we set up observations to explore different time scales able to identify day-weeks transients like short GRB afterglows and kilonovae, and slower evolving transients like supernovae or off-axis GRBs (cf. Tables~\ref{data} and \ref{data2}).

To prepare the Observing Blocks  (OBs)
we used a dedicated script named \emph{GWsky}.
\emph{GWsky} is a {\tt python}\footnote{http://www.python.org} tool devoted to effectively tile the sky
localization of a gravitational wave signal and provide accurate 
sequences of pointings optimized for each telescope\footnote{ \emph{GWsky} has
a Graphical User Interface optimized for fast and interactive 
telescope pointing operations. The field-of-view footprints are displayed 
in real time in the Aladin Sky Atlas via Simple Application Messaging
Protocol (SAMP) interoperability.} (Greco et al. in preparation). 
To define the sequence of pointings, \emph{GWsky} supplies information and descriptive statistics about telescope visibility, GW localization probability, presence of reference images and galaxies for each FoV footprint.

\begin{figure*}
\includegraphics[angle=0,width=\textwidth]{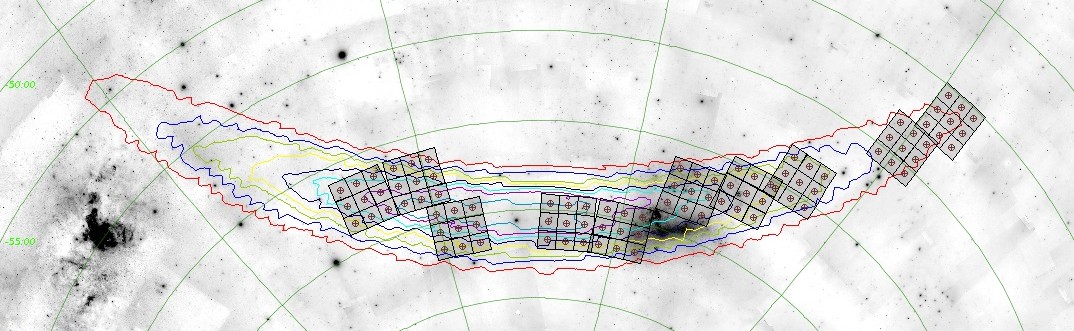}
\caption{Footprints of the VST $r$ band observations over the contours of the initially distributed cWB localization map of GW150914. Each square represents the VST Observing Block of 3$\times$3 deg$^2$. The lines represent the enclosed probabilities from a 90\% confidence level to a 10\% confidence level in steps of 10\%. The probability region localized in the northern hemisphere is not shown. The ten tiles enclose a localization probability of $\sim$ 29\%.  DSS--red image is shown in the background. An interactive skymap can be found in \url{https://www.grawita.inaf.it/highlights/}.
\label{fig:fig1}}
\end{figure*}

\begin{figure*}
\centering
\includegraphics[angle=0,width=0.45\textwidth]{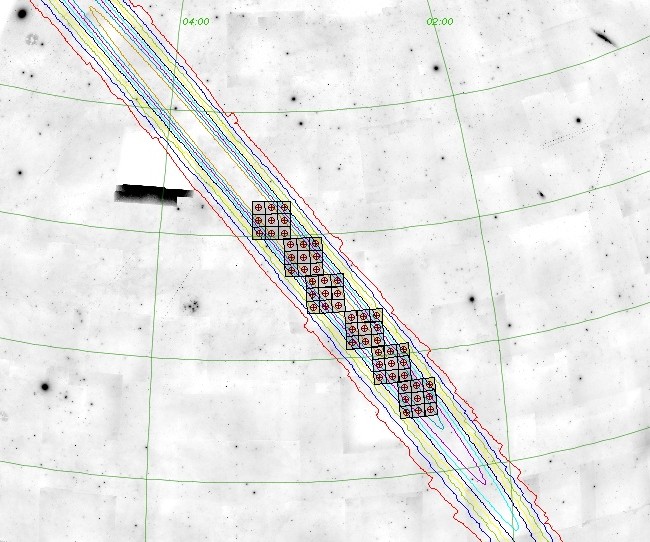}
\includegraphics[angle=0,width=0.45\textwidth]{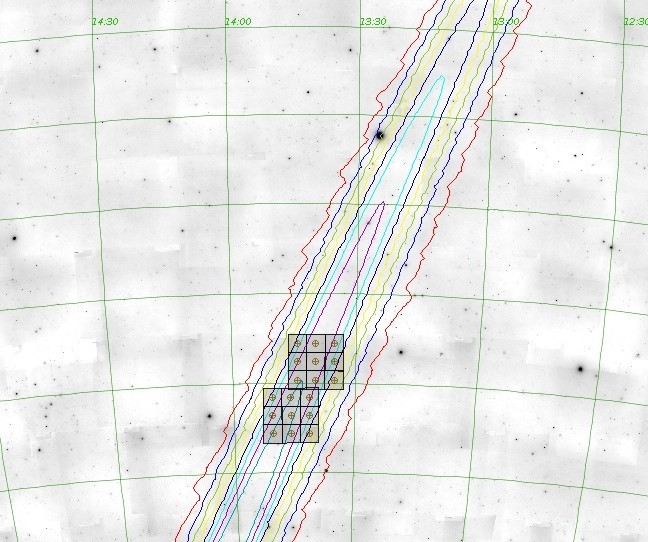}
\caption{Footprints of the VST $r$ band observations over the contours of the initially distributed BAYESTAR localization map of GW151226. From left to right, the VST coverage in the northern and  southern hemispheres is shown. Each square represents the VST Observing Block of 3$\times$3 deg$^2$. The lines represent the enclosed probabilities from a 90\% confidence level to a 10\% confidence level in steps of 10\%. The eight tiles enclose a localization probability of $\sim$ 9\%.  DSS--red image is shown in the background. An interactive skymap can be found in \url{https://www.grawita.inaf.it/highlights/}.
\label{fig:fig2}}
\end{figure*}

The sequence of the VST pointings for both GW events was defined optimizing the telescope visibility and maximizing the contained sky map probability accessible to the Paranal site, and excluding fields with bright objects and/or too crowded.
The typical VST OB contains groups of nine pointings (tiles) covering an area of 
$3 \times 3\; \rm{deg}^2$. For  each pointing, we obtained two exposures of 40\,s each dithered by $\sim 0.7-1.4$ arcmin. By doing this, the gaps in the OmegaCAM CCD mosaic are covered and most of the bad pixels and spurious events as cosmic rays are removed. The surveys of both events were performed in the $r$ band filter. Summary of the VST follow-ups of GW 150914 and 151226 are reported in Tab.\,\ref{data} and\,\ref{data2}, respectively.

\subsubsection*{GW150914}
The VST responded promptly to the GW150914 alert by executing six different OBs on 17th of September, $23$ hours after the alert and $2.9$ days after the binary black hole merger time \citep{GCN18336}. In this first night observations covered $54$ deg$^2$, corresponding approximately to the most probable region of the GW signal visible by VST having an airmass smaller than $2.5$.  The projected central region of the Large Magellanic Cloud (with a stellar density too high for our transient search) and the fields with bright objects were excluded from the observation. On 18th of September the sky map coverage was extended by adding a new set of four OBs, for a total coverage $90$ deg$^2$.
New monitoring of the $90$ deg$^2$  region was repeated \citep{GCN18397} over two months  for a total of six observation epochs.

Fig.\,\ref{fig:fig1} shows the cWB sky locations of GW 150914 and the VST FoV footprints superimposed on the DSS-red image. The coloured lines represent the enclosed probabilities from a 90\% confidence level to a 10\% confidence level in step of 10\%. For clarity, the probability region localized in the northern hemisphere is not shown. The VST observations captured a containment probability of 29\%. This value dropped to 10\% considering the LALinference sky map, which was shared with observers on 2016 January 13 \citep[][]{GCN18858}. This sky map generated using Bayesian Markov-chain Monte Carlo \citep{berry15}, modeling the in-spiral and merger phase and taking into account the calibration uncertainty is considered the most reliable and covers a 90\% credible region of $630$ deg$^2$ \citep[LALInf,][]{Abbott2016z}.

\subsubsection*{GW151226}
Also the response to GW151226 was rapid, 7.6 hours after the alert and 1.9 days after the merger event \citep{GCN18734}. Eight OBs covered $72$ deg$^2$ corresponding to the most probable region of the GW signal visible by VST and with an airmass smaller than $2.5$. Like for GW150914, the GW151226 survey consists of 6 epochs, spanning over one and a half month.

The two panels in Fig.\,\ref{fig:fig2} show the sequence of the VST pointings distributed across the BAYESTAR sky localization of GW151226 superimposed on the DSS-red image. The GW localization probability is concentrated in two long, thin arcs. 
Taking into account the characteristic ring-shaped region, the sequence of pointings runs along the inter-cardinal directions to maximize the integrated probability in each exposure. 
The VST observations captured a containment probability of 9\% of the initial BAYESTAR sky map and 7\% of the LALinference sky map, which was shared on January 18 \citep{GCN18889} and covers a 90\% credible region of $1240$ deg$^2$.

\section{Data Processing}
\label{dataproc}

\subsection{Pre-reduction}

Immediately after acquisition, the images are mirrored to ESO data archive, and then  transfered by an automatic procedure from ESO Headquarters to the VST Data Center in Naples. The first part of the image processing was performed using {\tt VST-tube}, which is the pipeline developed for  the VST-OmegaCAM mosaics \citep{Grado12}.  It includes pre-reduction, astrometric and photometric calibration and mosaic production.

Images are treated to remove instrumental signatures namely, applying overscan, correcting bias and flat-field, as well as performing gain equalization of the 32 CCDs and illumination correction. The astrometric calibration is obtained using both  positional information from overlapping sources and with reference to the 2MASS catalog.  The absolute photometric calibration is obtained using equatorial photometric standard star fields observed during the night and comparing the star measured magnitude with the SDSS catalogue\footnote{http://www.sdss.org}. A proper photometric calibration is evaluated using the {\tt Photcal} tool \citep{radovich} for each night. The relative photometric calibration of the images is obtained minimizing the quadratic sum of differences in magnitude between sources in overlapping observations. The tool used for both the astrometric and photometric calibration tasks is {\tt SCAMP} \citep{BertinSCAMP}. 
Finally the images are re-sampled and combined to create a stacked mosaic for each pointing. In order to simplify the subsequent image subtraction analysis, for each pointing the mosaics at the different epochs are registered and aligned to the same pixel grid. In this way, each pixel in the mosaic frame corresponds to the same sky coordinates for all the epochs. For further details on the data reduction see \cite{capaccioli2015}. 

With the current hardware, the time needed to process one epoch of data of the VST follow-up campaigns here described, including the production of the {\em SExtractor} \citep{BertinSEX} catalogs and all the quality control checks, amounts to about 6 hours.

\subsection{Transient search}

In order to search for variable and transient sources, the images were analysed by using two independent procedures. One is based on the comparison of the photometric measurements of all the sources in the VST field obtained at different epochs. The second is based on the analysis of the difference of images
following the approach of the supernova (SN) search program recently completed with the VST \citep{Botticella2016}. 

The two approaches are intended to be complementary, with the first typically more rapid and less sensitive to image defects and the latter more effective for sources projected over extended objects or in case of strong crowding. In the following, we report some details about both the data analysis approaches. Taking into account the largely unknown properties of the possible EM gravitational wave counterpart we decided to not use model-based priors in the candidate selection.
For both procedures, the main goal of our analysis is to identify sources showing a ``significant" brightness variation, either raising  or declining flux, during the period of monitoring, that can be associated to extra-galactic events.

\subsubsection{The photometric pipeline ({\tt ph-pipe})}

The photometric pipeline is intended to provide a list of ``interesting" transients in low-latency to organise immediate follow-up activities. The computation time can be particularly rapid, e.g. just a few minutes for each epoch VST surveyed area. The weakness of this approach is that sources closer than about a Point Spread Function (PSF) size or embedded in extended objects can be difficult to detect and therefore can possibly remain unidentified.

The procedure has been coded in the {\tt python}
~(version 3.5.1) language making use of libraries part of the {\tt anaconda}\footnote{https://docs.continuum.io/anaconda/index}~(version 2.4.1) distribution.  The procedure includes a number of basic tools to manage the datasets, i.e. source extraction, classification, information retrieval, mathematical operations, visualization, etc. Data are stored and managed as {\tt astropy}\footnote{http://www.astropy.org}~(version 1.2.1) tables. 

The analysis is based on the following steps:

\begin{enumerate}

\item The 
{\tt SExtractor} package (Bertin \& Arnouts 1996), as implemented in the python module {\tt sep}\footnote{https://sep.readthedocs.org/en/v0.5.x/} (version 0.5.2), was used for source extraction. This algorithm gives the best results considering the request of a rapid running time. The extraction threshold is set at 5$\sigma$. 

\item Each source list is then cleaned removing obvious artifacts by checking various shape parameters (roundness, full width at half maximum, etc.). Then a quality flag based on the ``weight" maps generated by the VST reduction procedure \citep{capaccioli2015} is attributed to the detected objects. All the sources are processed but only those associated to the best exposed frame zones 
are used to tune the statistical analyses (described below) aimed at identifying transients or variable objects.

\item Aperture photometry is measured for all the sources at each epoch.  Although at the expense of longer computation time, the more reliable algorithm {\tt DAOPHOT} \citep{Stetson87}, as coded in the {\tt PythonPhot}\footnote{https://github.com/djones1040/PythonPhot} (version 1.0.dev) module, is used rather then other quicker alternatives. The magnitudes at each epoch are normalised to those of the reference epoch, typically but not necessarily the first in chronological order, computing the median difference of the magnitudes of objects with the highest quality flag. Finally, the angular distance and the magnitude difference from the closest neighbors are computed for each source to evaluate the crowding.

\item The source list is cross-correlated (0.5\arcsec\ radius) with the Initial GAIA source list \citep[IGSL, ][]{SmaNic14} and later, when it became available, with the GAIA catalogue (DR1 release)\footnote{http://vizier.u-strasbg.fr/viz-bin/VizieR?-source=I/337}, saving the uncatalogued sources and sources catalogued as extended (possible GW host galaxies) for further analysis.
 This typically removes about 40\% of the detected objects, depending on the depth of the observations and the Galactic coordinates of the observed field. The risk of erroneously remove the nucleus of some faint or far galaxy, wrongly classified in these catalogs as point-like sources, is of course present. We checked that within the magnitude limits of the considered catalogs (and considering the distance range of the counterparts to GW events we are looking for) most of the extended objects are indeed correctly identified and classified. The SDSS\footnote{http://www.sdss.org} and the Pan-STARRS\footnote{http://panstarrs.stsci.edu} data releases are also used in case the analysed areas are covered by these surveys.

\item A ``merit function" is derived taking into account several parameters as variability indices (i.e. maximum-minimum magnitude, $\chi^2$ of a constant magnitude fit, proximity to extended objects, signal-to-noise ratio, crowding). The higher the value of the merit function the more interesting the variability of the transient object is.

\item The selection of the interesting objects, i.e. those showing a large variability and those with the higher merit (the merit also includes variability information although not necessarily large variability implies a high merit), including objects previously undetected or disappeared during the monitoring, is a multi-step process. First of all, the highest quality ranked objects are binned in magnitude to compute the sigma-clipped averages and the standard deviations of the magnitude difference for each available epoch. Then, all the objects showing variability larger than a given threshold (e.g. 5-7 $\sigma$, in our cases) between at least two epochs are selected (this practically corresponds to a magnitude difference larger than about 0.5\,mag for good quality photometric information). The whole procedure is affected by some fraction of false positives due to inaccuracies of the derived photometry for sources with bright close companions since a seeing variation among the analyzed epochs can induce a spurious magnitude variation. \label{select}
 
\item The list of (highly) variable objects is cross-correlated (2\arcsec\ radius) with the SIMBAD astronomical database \citep{Wenger2000} to identify already classified sources and with the list of minor planets provided by the SkyBot\footnote{http://vo.imcce.fr/webservices/skybot/} portal at the epoch of observation. This piece of information is stored but the cross-correlated objects are not removed from the list yet.
 
\item The last step of the analysis consists in the computation of PSF photometry for the selected objects
again using {\tt PythonPhot} module. The PSF is derived selecting automatically at least 10 isolated stars in a suitable magnitude range. In order to keep the computation time within acceptable limits, PSF photometry is derived only for the objects of interest without carrying out a simultaneous fit of the sources in the area of the target of interest. For moderate crowding this is already sufficient to derive reliable photometric information even in case of large seeing variation.

\item Then, by means of the PSF photometry, step\,\ref{select} is repeated and the list of objects surviving the automatic selection is sent to a repository for a further final check via visual inspection. Stamps of these objects for each epoch are produced to aid the visual inspection and FITS files of any size around them can also be produced if needed. It is also possible to produce light-curves, to convert the list of candidates to formats suited for various graphical tools (e.g. the starlink GAIA FITS viewer\footnote{http://star-www.dur.ac.uk/~pdraper/gaia/gaia.html}).

\end{enumerate}

As an example, for the observations taken after the GW150914 event the number of extracted sources ranged from a few tens of thousands in high Galactic latitude fields, to about half a million for fields nearby the Large Magellanic Cloud. About three million sources per each epoch of our monitoring and a total of about nine million of sources were extracted and analysed. 
The number of highly variable objects, satisfying our selection criteria and not present in the GAIA catalog, resulted to be 54239, about 0.6\% of the initial list.
Choosing only the sources with higher score we remain with about 5000 candidates. The last cleaning is carried out by visual check, candidates affected by obvious photometric errors due to crowding, faintness, or image defects are removed. Candidates showing good quality light-curves that can be classified basing on known variable class templates (RR Lyare, Cepheids, etc.) are also removed form the list, this step indeed allows us to clean the majority of the remaining candidates. Finally, candidates showing light-curves grossly consistent with the expectations for explosive phenomena as GRB afterglows, SNae and macronovae, or candidates laying nearby extended objects (i.e. galaxies) are saved for further processing defining a final list of 939 sources (cf. Sect.~\ref{G15}).

\subsubsection{The image difference pipeline ({\tt diff-pipe})}

A widely used, most effective approach for transient detection is based on the difference of images taken at different epochs. To implement this approach for the survey described in this paper we developed a dedicated pipeline exploiting our experience with the medium-redshift SN search done with the VST \citep[SUDARE project,][]{Cappellaro15}. The pipeline is a collection of {\tt python} scripts that include specialized tools for data analysis, e.g. {\tt SExtractor}\footnote{http://www.astromatic.net/software/sextractor} \citep{BertinSEX} for source extraction and {\tt topcat}\footnote{http://www.star.bris.ac.uk/~mbt/topcat/}/{\tt stilts}\footnote{http://www.star.bris.ac.uk/~mbt/stilts/} for catalog handling. For optical images taken from the ground, a main problem is that the PSF is different at different epochs, due to the variable seeing. The PSF match is secured by the {\tt hotpants}\footnote{http://www.astro.washington.edu/users/becker /v2.0/hotpants.html} code \citep{Becker2015}, an implementation of the \citet{Alard99} algorithm for image analysis.

The analysis is based on the following steps:

\begin{enumerate}
\item For each image the VSTtube \citep{Grado12} pipeline produces a bad pixels mask with specific flags. The areas enclosing bright/saturated stars, that leave spurious residuals in the image difference, are also masked.
	
\item  We compute the difference of images taken at different epochs. For PSF match, by comparing sources in common between the two images, the image with the best seeing is degraded to match the other image.
 In an ideal case one would like to use template images taken before the actual search epochs. Unfortunately, such templates are not available for the specific area monitored in our survey and therefore we used as template the image taken at the latest epochs.
With this approach we are able to detect as positive sources in the difference image all the transients that 
at the latest epoch disappeared or, in general, are fainter than in the previous epochs.  
On the contrary, sources that are brighter at the latest epoch leave a negative residual in the difference image and would not be detected. The latter ones can be detected by searching the ``negative" difference image  that is obtained by multiplying the regular difference by $-1$ (see next).
	
\item{\tt SExtractor} is used to detect positive sources in the difference image (transient candidates). 
We also search for negative differences to guarantee completeness for raising  or declining transients.
The number of detected sources strongly depends on the adopted threshold, defined in unit of the background noise. In this experiment we use a  $1.5\,\sigma$ threshold.
From the list of detected sources  we delete all sources occurring in a flagged area of the masked image.
		
\item The list of candidates contains a large number of spurious objects that can be related to small mis-alignment of the images, improper flux scalings, incorrect PSF convolution or to not well masked CCD defects and cosmic rays.

To filter out the spurious candidates, we use a ranking approach. To each candidate we assign 
an initial score that is decreased/increased depending on different source parameters either provided
by SExtractor or measured directly on the difference image. By using a combination of different parameters,  
we test whether the source detected in the difference image is consistent with being a genuine stellar source.
The ranking scores are calibrated by means of artificial star experiments to ensure that good candidates obtain a positive score.

The main SExtractor parameters used to derive the ranking  for each candidates are: {\rm FWHM, ISOAREA, FLUX\_RADIUS} and {\rm CLASS\_STAR}.
In addition, we penalized transient candidates very close to a bright star of the reference image and/or  those for which the ratio of positive/negative pixels in the defined aperture is below a specific threshold. In fact, in many cases small PSF variations produce positive/negative pairs in the difference image.

In this scheme, we also allow for positive attributes intended to promote specific type of sources.  
In particular, we promote transients found near galaxies with the idea that these are worth a second look.

\item The catalogs of sources detected at different epochs in each pointing are merged. In this final catalog we include only candidates with scores above a selected score threshold, though we also record the number of independent detections for each candidate regardless of the score.

\item We cross check our candidate list with the {\em SIMBAD} database using a search radius of 2\arcsec\ with the purpose to identify known variable sources. While we do not expect them to be the EM counterpart, known sources are useful to test the pipeline performance.

\item For each candidate we produce a stamp for visual inspection including the portion of the original images at the different epochs along with the same area in the respective difference images. If needed, one can also produce stamps for specific coordinates, not corresponding to detected transients. This is useful to check for candidates detected by other searches.

\item Finally, we perform detailed artificial star experiments with the aim to measure the search efficiency as a function of magnitude and provide rates or, in case, upper limits for specific kind of transients.
\end{enumerate}

As an example, for the case of GW150914, the procedure produced a list of about 170000 transient candidates (with an adopted threshold of $1.5\sigma$ of the background noise) many with multiple detections. The scoring algorithm reduces this number by one order of magnitude: the final list includes 33787 distinct candidates of which 11271 candidates with high score that are taken as bona-fide genuine transients. 
Finally, we performed a visual inspection  concluding that $\sim 30\%$ are obvious false positive, not recognized by the ranking algorithm.

The image difference pipeline was definitely more time consuming than the photometric pipeline: e.g. the computing time for the typical case (90 deg$^2$, at six epochs) was around 2 days, that is fairly long for low-latency search. For future triggers we have implemented parallel version of the pipeline, using the {\tt python} modulus {\tt pp}\footnote{https://github.com/uqfoundation/ppft}. This will reduce the required time by a factor $\sim 5$.

A comparison between the transients identified by the two pipelines shows that, as expected, the image-difference pipeline is more effective, in particular for objects very close to extended sources. However, the photometric pipeline is less affected by image defects as halos of very bright or saturated stars, offering a profitable synergy. Typically, a percentage ranging from 80 to 90\% of the transients identified with the photometric pipeline are also recorded by the image-difference pipeline.

\subsubsection{The detection efficiency}

In order to measure our search performance and to tune the observing strategy, we performed extensive artificial star experiments. To this aim we use the {\tt daophot} package to derive the PSF for each of the searched image and then we add a number of artificial stars of different magnitudes in random positions. Then, we run the image difference pipeline and count the number of artificial stars that are recovered with a score above the adopted threshold. The ratio of recovered over injected stars gives the detection efficiency as a function of magnitude. 
An example of the outcome of this procedure is shown in Fig.~\ref{artstar1} for three different pointings following the GW151226 trigger. The detection efficiency vs. magnitude empirical relation is well fitted by a simple function \citep{Cappellaro15} and can be used to measure the parameter DE$_{50}$,
defined as the magnitude at which the detection efficiency drops  to 50\% of the maximum value. This depends first of all on sky conditions, transparency and seeing, but also on field specific properties, in particular crowdedness and contamination by bright stars.
In Fig.~\ref{artstar2} we show the measurements of DE$_{50}$ for all the pointings of the two GW triggers as a function of seeing. We notice that, for good sky conditions our survey can detect transients down to $r \sim22$ though most observations are in the range $20-22$ mag. On the other hand, in case of poor seeing (FWHM$>1.5$ arcsec) the magnitude limit is $\sim 20$ mag. 

\begin{figure}
\begin{center}
\includegraphics[width=0.48\textwidth]{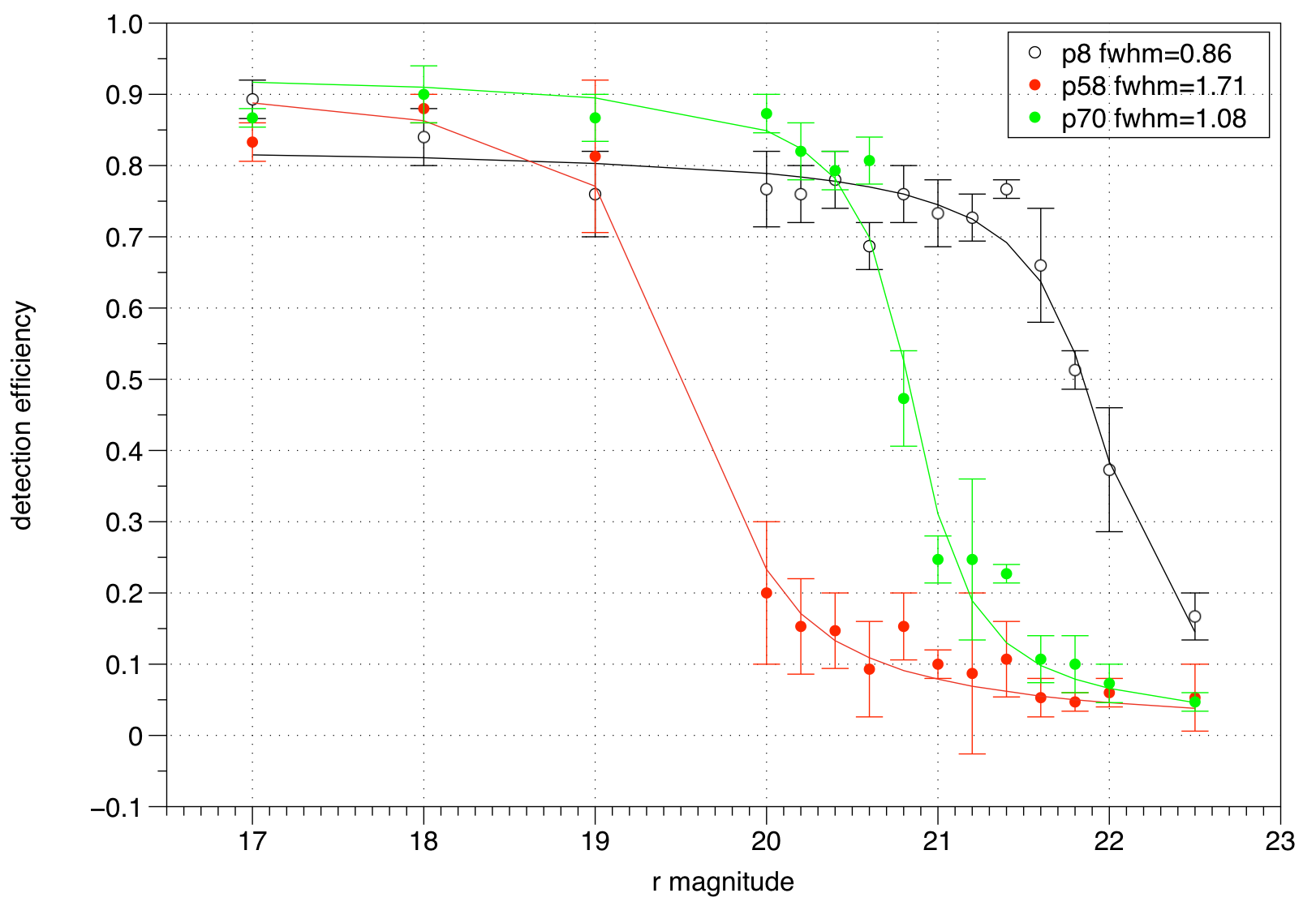}
\caption{Example of the output of artificial star experiments. The detection efficiency  (DE) is defined as the ratio between the number of detected stars and the number of injected stars in specific magnitude. The plot shows the correlation between DE and the magnitude for three pointings of GW151226 (p8, p58, p70). 
}\label{artstar1}
\end{center}
\end{figure}

\begin{figure}
\begin{center}
\includegraphics[width=0.48\textwidth]{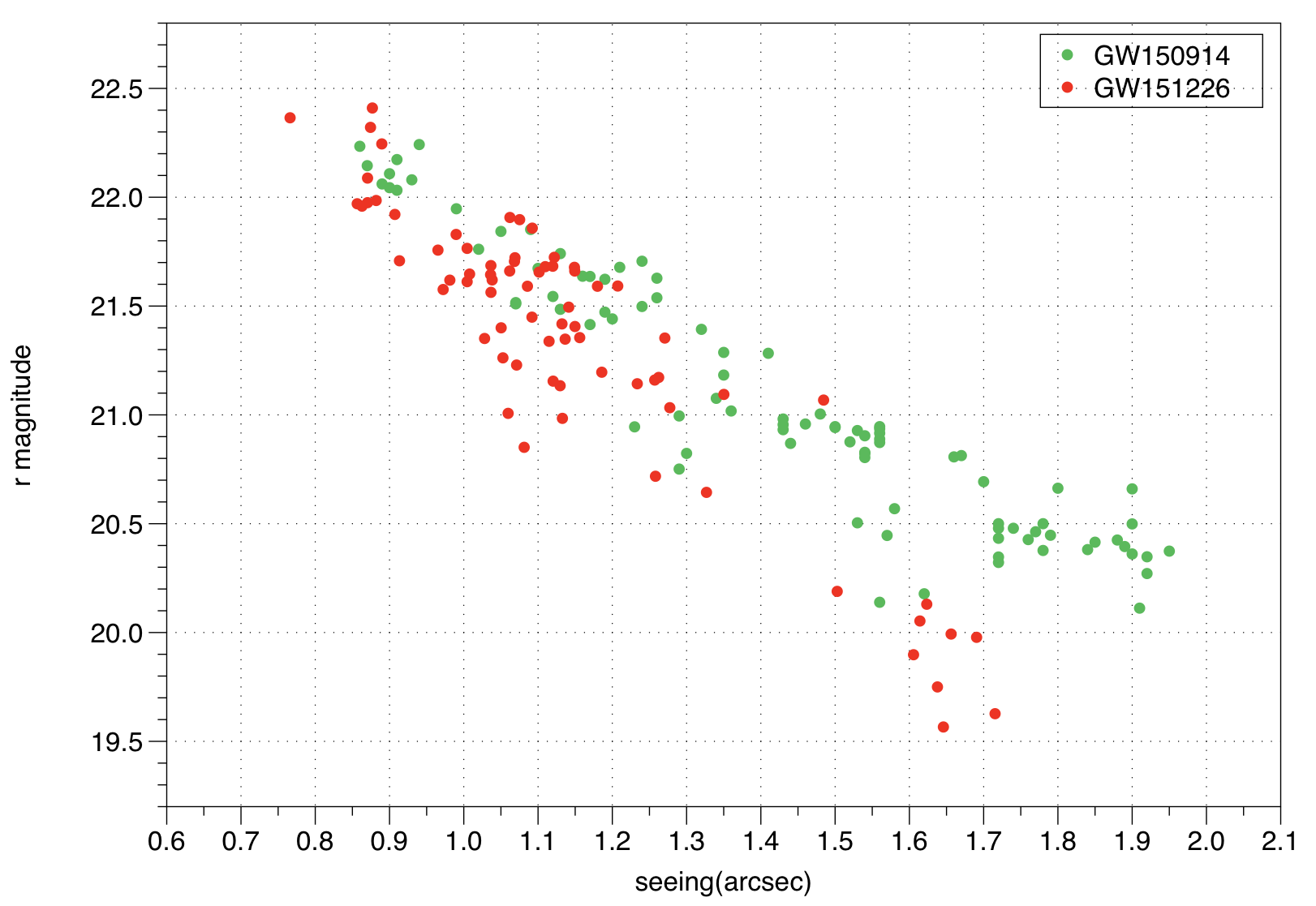}
\caption{The limiting magnitude for transient detection (DE$_{50}$) as a function of seeing for the pointings of the two triggers discussed in this paper. The scatter is due to the fact that other factors are affecting the DE, first of all sky transparency.
}\label{artstar2}
\end{center}
\end{figure}

\section{Results}
\label{result-foll}

We now know that both the gravitational wave events considered here, GW150914 and GW151226, were generated by coalescence of black-holes. In the current scenario strong electromagnetic radiation is not expected to occur, and in fact none of the transients identified by the worldwide astronomical effort could be linked to the observed GW events. However, the analysis of the data obtained in response to the GW triggers is important both for evaluating the search performances and for tuning future counterpart searches. 
In the following we will give an overview of the results of our search and describe a few representative transients, typically candidate SNe, detected by our analyses with the purpose to illustrate pros and cons of our approach.

An important limitation for our analysis is that the sky areas surveyed after the two triggers were never observed before with the VST telescope and therefore we do not have access to proper reference images. The consequence is that for an efficient transient search we had to wait for the completion of the monitoring campaign and could not activate immediate follow up. For this reason, we only have few cases of candidate SNe associated with galaxies with known redshift, for which we propose a possible classification.

Finally, for an external check of our survey performances, we compared the candidate detected by our pipelines with those found by other searches, when available.

\subsection{GW150914}\label{G15}

As described in Section~\ref{obs}, the VST observations started 2.9 days after the occurrence of the GW150914 event and just 1 day after the alert. The 90 deg$^2$ observed sky area captured 29\% of the initial cWB sky map probability and 10\% of the more accurate LALInference sky map. Indeed, this latter sky map is more suitable for BBH mergers but it was made available only on January 2016, when most of the EM follow-ups on GW150914 were already over. 
Prompt response, survey area and depth make a unique combination of features of our VST survey (see Fig.~\ref{VSTperf}) matched only by the DECam survey \citep{DECam} at least for what concerns the combination of depth and area of the survey.

\begin{figure}
\center
\includegraphics[scale=0.3]{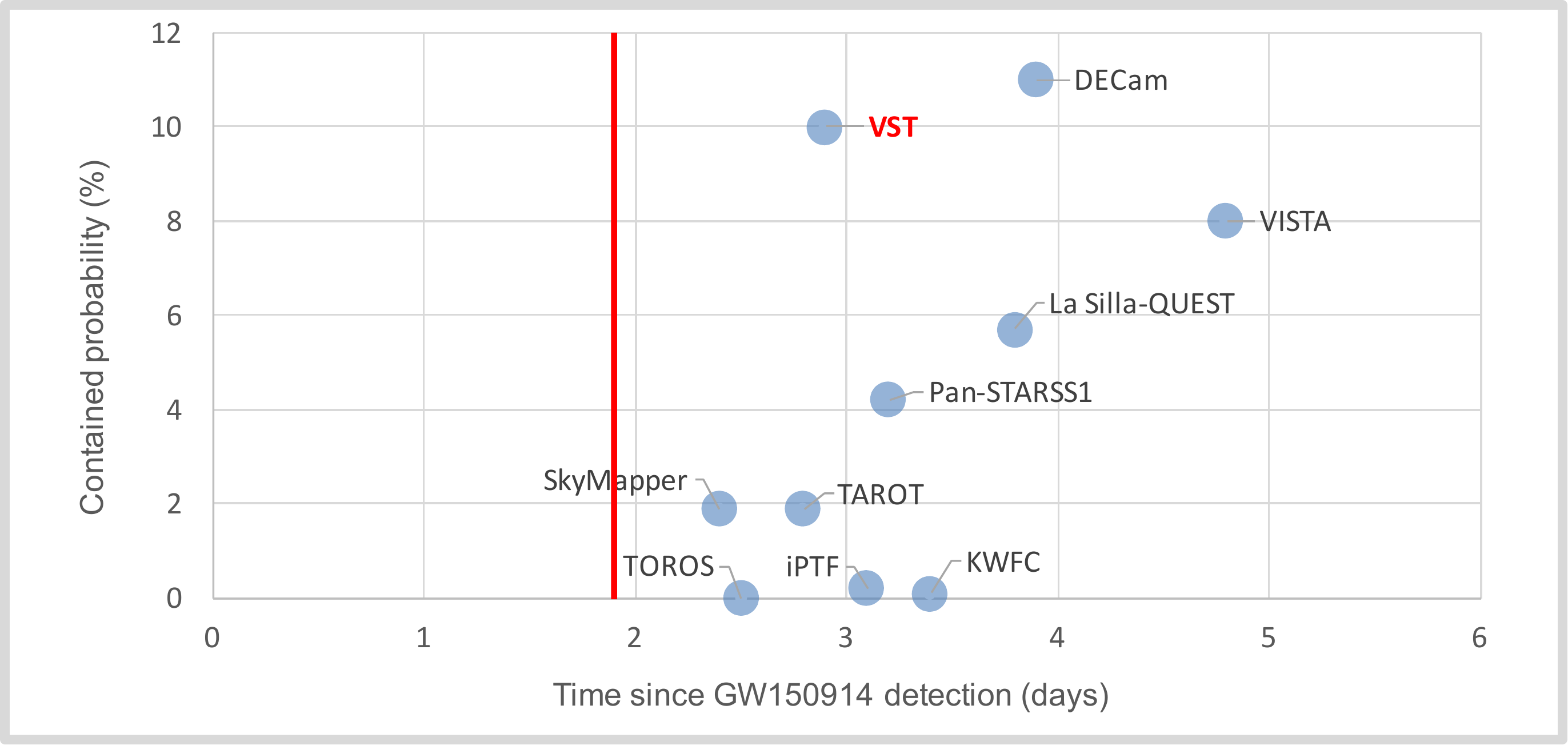}
\includegraphics[scale=0.3]{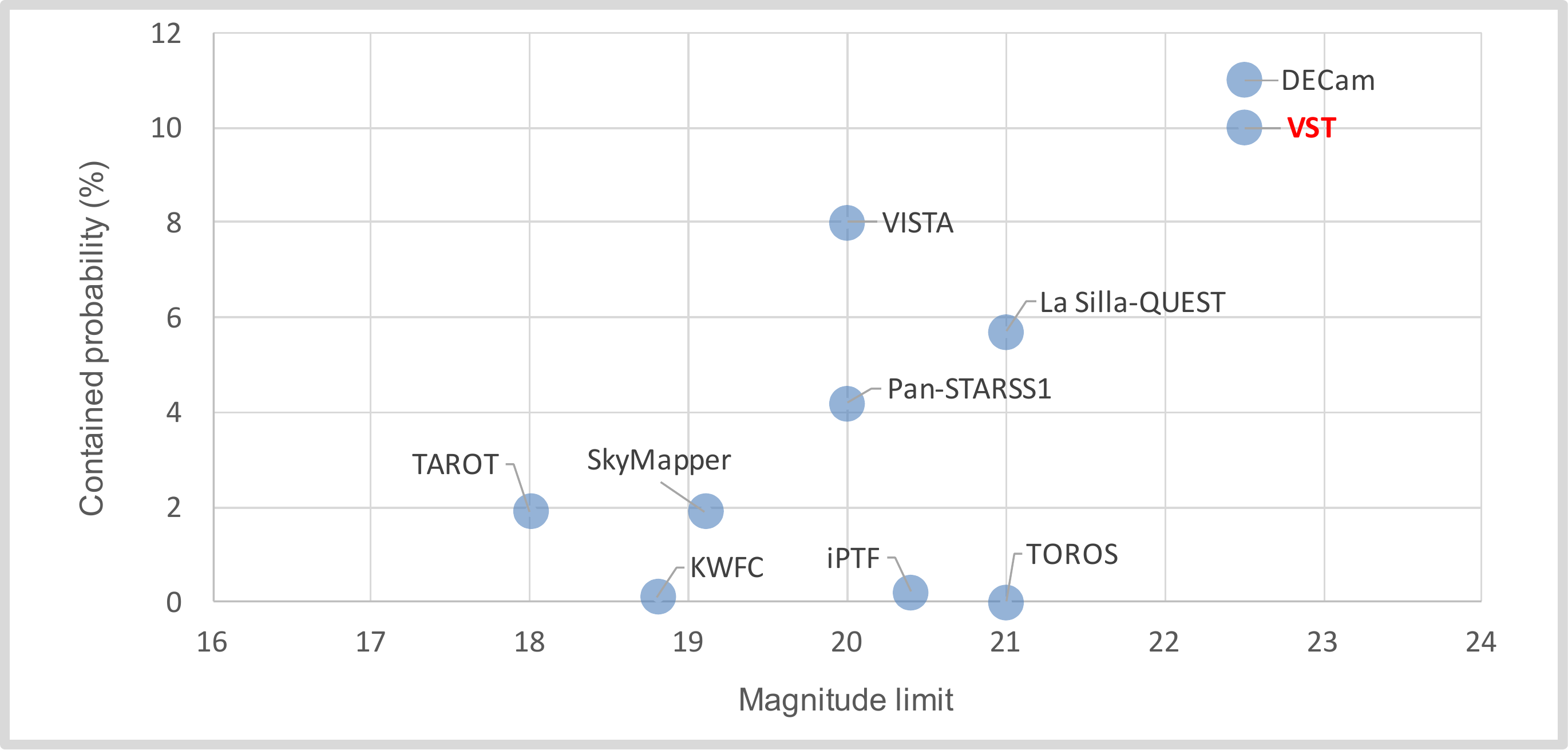}
\caption{VST performance. In the top panel the time response of VST in terms of time 
and contained probability is compared to other facilities. The red vertical line marks the 
time of the LVC alert to the astronomical community.  A similar comparison is plotted
in the lower panel but in the abscissa the approximate magnitude limits are reported. The magnitude limits 
refer to different photometric bands.
The data are from \citet[][]{Abbott2016f,Abbott2016z}.}
\label{VSTperf}
\end{figure}

The total list of variable/transient objects selected by the {\tt diff-pipe} consists of 33787 sources (of which 11271 with high score). The number of sources provided by the {\tt ph-pipe} is 939. More than 90\% of them are also detected by the {\tt diff-pipe}. The smaller number of sources detected by the {\tt ph-pipe} is due to $i)$ the removal of all the ``bright" and/or previously known variable sources after the match with the GAIA catalog and $ii)$ the much higher adopted detection threshold. Most of the sources identified by the {\tt ph-pipe} and not included in the catalog produced by the {\tt diff-pipe} turned out to be real and were typically located in regions that needed to be masked for a reliable image subtraction.
Many of the {\tt diff-pipe} candidates are known variables. As a further text, we applied the same selection criteria of the {\tt ph-pipe} to the list of the 33787 variable/transient sources identified by {\tt diff-pipe}. The selection produces a list of about 3000 objects. This last sample still includes known variable sources (more than 400) or objects whose light-curves can be classified with known templates, or possible defects in the subtraction procedure. As expected, the {\tt diff-pipe} is more effective in finding variable/transient objects than the {\tt ph-pipe}, although the final cleaned lists also contain objects that are found by one pipeline only.

\begin{table}
\centering
\caption{Number of variable and total detected sources ({\tt diff-pipe}) within the $3 \times 3\; \rm{deg}^2$ areas covered by each of the 9 tiled observations. Those close to the LMC are clearly recognizable by the large number of sources.}
\label{tab:sourceratio}
\begin{tabular}{rrrr}
\hline
\hline
\multicolumn{1}{c}{RA} &
\multicolumn{1}{c}{Dec} & Num. var &  Tot. sources  \\
\multicolumn{1}{c}{J2000} & \multicolumn{1}{c}{J2000} &  &  \\
\hline
 58.208846  &  $-56.949515$  &   196  &    34345 \\  
 60.652964  &  $-59.855304$  &   430  &    36057 \\  
 68.948300   &  $-64.802918$  &   645  &    69077 \\ 
 74.729746   &  $-66.793713$  &  6225  &   676621 \\ 
 82.166543   &  $-67.952724$  & 14590  &  1083748 \\  
 91.163807   &  $-71.180392$  &  6337  &   720924 \\  
 100.348601  &  $-71.180473$  &  1923  &   147827 \\  
 118.562044  &  $-71.090518$  &   654  &    98150 \\  
 122.909379  &  $-67.971038$  &   700  &   125286 \\  
 131.090822  &  $-67.972011$  &  2087  &   183930 \\  
\hline
\end{tabular}
\end{table}

As it can be seen from Fig.~\ref{fig:fig1}, some of the VST fields overlap with the outskirt of the Large Magellanic Cloud (LMC) which contributes with a large number of relatively bright stars and many variable sources. This effect is clearly visible from the statistics of detected and variable sources in the fields as reported in Table~\ref{tab:sourceratio}. This represents a severe contamination problem in the search for the possible GW counterpart. However, the LMC has been the target of a very successful monitoring campaign by the Optical Gravitational Lensing Experiment (OGLE)\footnote{http://ogle.astrouw.edu.pl}. The OGLE survey is fairly complete down to mag $\sim 20$ and has already identified many of the variable stars in the field.
A cross-check of our {\tt diff-pipe} candidate catalog against the SIMBAD database gave a match for 6722 objects of which 6309 identified with different type of variable sources, mainly RRLyrae (48\%), eclipsing  binaries (23\%) and a good number of Long Period Variables, semi-regular and Mira (23\%). The sky distribution of the matched sources reflects the LMC coverage by both our and the OGLE surveys. We notice that, as appropriate, the fraction of SIMBAD variable sources identified among our high score transient candidates is much higher (55\%) than for the low score candidates (26\%).

\begin{table*}
\begin{center}
\caption{Coordinates of the known or newly identified sources (SNe or candidate SNe) derived from the GW\,150914 follow-up campaign discussed in this section.}
\label{tab:coords}.
\hspace*{-0.5cm}
\begin{tabular}{lllll}
\hline
\hline
\multicolumn{1}{c}{Id} & \multicolumn{1}{c}{RA} & \multicolumn{1}{c}{Dec} & Alternate Id & Note \\
     & \multicolumn{1}{c}{J2000} & \multicolumn{1}{c}{J2000} & \\
\hline
VSTJ54.55560-57.56763 & 3:38:13.34 & -57:34:03.5 & & SN candidate \\ 
VSTJ56.28055-57.91392 & 3:45:07.33 & -57:54:50.1 & & SN candidate \\
VSTJ57.77559-59.13990 & 3:51:06.14 & -59:08:23.6 & & SN\,Ia or Ib/c candidate, $z \sim 0.11$ \\
VSTJ60.54727-59.91890 & 4:02:11.34 & -59:55:08.0 & & SN candidate \\
VSTJ61.20106-59.98816 & 4:04:48.25 & -59:59:17.4 & & SN candidate \\
VSTJ69.10694-62.79775 & 4:36:25.67 & -62:47:51.9 & OGLE15oa & SN Ia \\
VSTJ69.55973-64.47081 & 4:38:14.34 & -64:28:14.9 & & SN candidate \\
VSTJ71.71864-65.89735 & 4:46:52.47 & -65:53:50.5 & OGLE-2014-SN-094 & AGN candidate  \\
VSTJ113.77187-69.13147 & 7:35:05.25 & -69:07:53.3 & SN\,2015J & SN IIn, $z \sim 0.0054$ \\
VSTJ114.06567-69.50639 & 7:36:15.76 & -69:30:23.0 & SN\,2015F & SN Ia, $z \sim 0.0048$ \\
VSTJ119.64230-66.71255 & 7:58:34.15 & -66:42:45.2 & & SN\,Ia or Ib/c candidate, $z \sim 0.047$ \\
\hline
\end{tabular}
\end{center}
\end{table*}

\subsubsection{Previously discovered Transients}

Searching the list of recent SNe\footnote{We used the update version of the Asiago SN catalog \citep[http://sngroup.oapd.inaf.it/asnc.html,][]{Barbon99}}, we found that in the time window of interest for our search, three SNe  and one SN candidate were reported that are expected to be visible in our search images,  All these sources were detected in our images, and in particular:

\begin{itemize}

\item 
SN\,2015F was discovered by LOSS in March 2015 \citep{Monard15} in NGC\,2442 ($z \sim 0.0048$) and classified as type Ia with an apparent magnitude at peak of $\sim17.4$. The object was detected by our pipeline in the radioactive declining tail.

\item 
SN\,2015J was discovered on 2015-01-16 \citep{Brown14,Scalzo15} and classified as type IIn at a redshift $z \sim 0.0054$ \citep{Guillochon17}. In our images it was still fairly bright at $r \sim 17.8$, fading to $r \sim 18.5$ in a month (Fig.\,\ref{fig:candsknown}, right panel).

\item 
OGLE15oa was discovered on 2015-10-16 (by OGLE-IV Real-time Transient Search, \citet{Wyrzykowski14}) and was classified as a type Ia about 20 days after maximum on 2015-11-09 by \citet{Dennefeld15}. Most of our images are pre-discovery and the pipeline detected the transient at mag $r \sim 18.8$ in the images obtained in the last epoch, 2015-11-16.

\item A special case is 
OGLE-2014-SN-094,
which was discovered on 2014-10-06 and initially announced as a SN candidate \citep{Wyrzykowski14}. The source showed a second outburst in May 2015 and again in Nov 2015 \citep{Guillochon17}. We detected the source at the end of our monitoring period at a magnitude similar to that at discovery ($r\sim 19.5$, Fig.\,\ref{fig:candsknown}, left panel). The photometric history indicates that this is not a SN but more likely an AGN. A UV bright source, 
GALEXMSC J044652.36-655349.9, 
was also detected at the same position\footnote{http://ned.ipac.caltech.edu}. 
\end{itemize}

\begin{figure*}
\begin{center}
\includegraphics[scale=0.38]{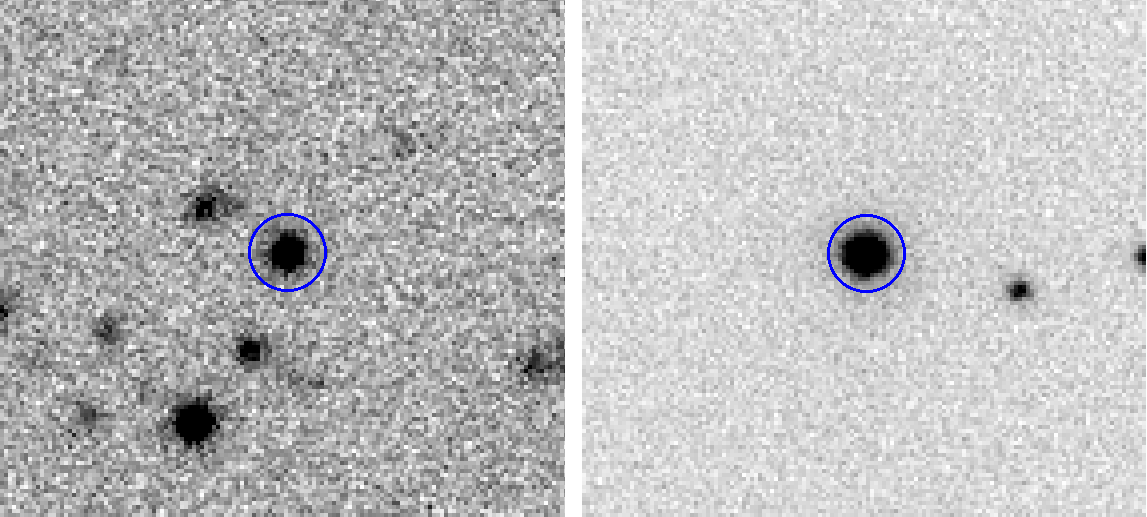}
\end{center}
\caption{\textit{Left:} The SN candidate 
OGLE-2014-SN-094 observed on 2015 Nov. 11. \textit{Right:} The SN\,IIn 
SN\,2015J at $z \sim 0.0054$ observed on 2015 Sept. 15. 
The blue annuli represent the position identified by our pipelines}\label{fig:candsknown}
\end{figure*}

\begin{figure*}
\begin{center}
\includegraphics[scale=0.21]{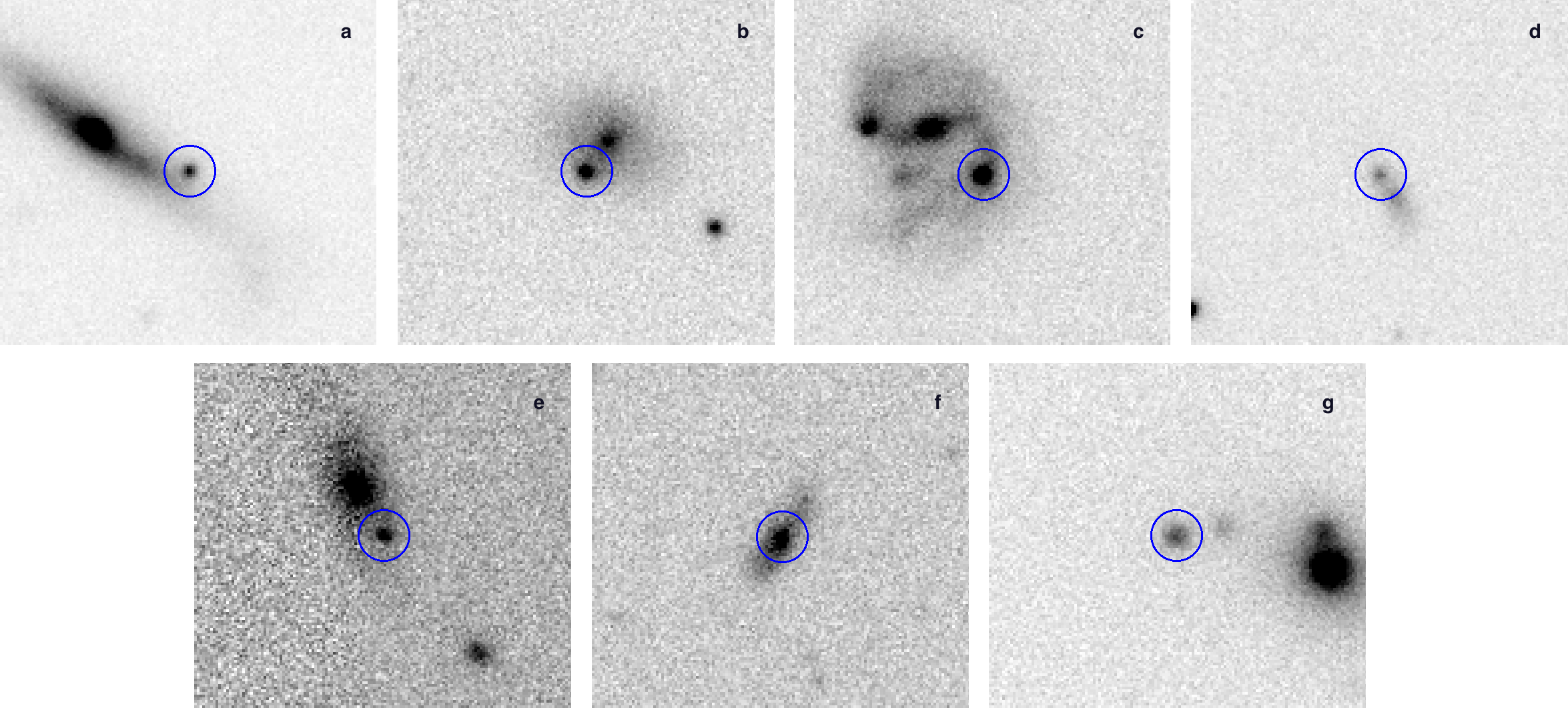}
\end{center}
\caption{SN candidates identified in our survey after GW150914. \textit{a}. VSTJ54.55560-57.56763 observed on 2015, Sept. 17. \textit{b}. VSTJ56.28055-57.91392 observed on 2015, Oct.13. \textit{c}. VSTJ57.77559-59.13990 observed observed on 2015, Sept. 18. The galaxy is at redshift $z \sim 0.11$. \textit{d}. VSTJ60.54735-59.91899 observed on 2015, Sept. 30. \textit{e}. VSTJ61.20106-59.98816 observed on 2015, Sept. 30.
\textit{f}. VSTJ69.55986-64.47089 observed on 2015, Sept. 17.
\textit{g}. VSTJ119.64244-66.71264 observed on 2015, Oct. 13.
In all images the showed field sizes are 30$\times$30 arcsec, North is up and East to the left.
The blue annuli represent the position identified by our pipelines.
}
\label{fig:candsunknown}
\end{figure*}

\subsubsection{Transient candidates}

In addition, we also singled out a few objects that most likely are previously undiscovered SNe (Fig.\,\ref{fig:candsunknown}). 

\begin{itemize}

\item VSTJ54.55560-57.56763: the source was fading after the  detection during our first epoch observation. It is located close to an edge-on spiral galaxy 
{PGC\,145743} \citep[HyperLEDA, ][]{Makarov14}. No redshift is available.

\item VSTJ56.28055-57.91392: this source was caught during brightening. It is located close to a spheroidal galaxy (
{2MASXJ03450711-5754466} in HyperLEDA). No redshift is available.

\item VSTJ57.77559-59.13990 was likely detected close to peak ($r \sim 19.4$~mag). It was located in the arm of the face-on, barred spiral galaxy 
PGC\,141969 at redshift $z\sim 0.11$ \citep[The 6dF Galaxy Survey Redshift Catalogue,][]{Jones09}. The transient absolute magnitude was then brighter than $\sim -19$. In Fig.\,\ref{fig:snz}, top panel, we show our photometry (assuming the distance obtained from the redshift of the likely host galaxy, i.e. $z \sim 0.11$) superposed 
to the 
light-curve of SN\,1998bw \citep{Galama98,Patat01,Iwamoto98}. SN\,1998bw was associated with the long GRB\,980425 \citep{Pian00} and it is the prototype of the   broad-lined stripped-envelope SNe events SN\,Ib/c \citep{Iwamoto98, Mazzali13}.
From this comparison we estimate that the SN explosion occurred about three weeks before our first observation, that is in late August 2015. Interestingly, the {\it Fermi}-GBM online archive\footnote{https://heasarc.gsfc.nasa.gov
} shows that on 2015 August 27 a GRB (burst time 18:50:12.969 UT, t$_{90} \sim 10$\,s, RA$_{J2000}$=04:33:12.0, DEC$_{J2000}$=-60:00:00) was detected at a distance of about 5.5$^\circ$, consistent within the error with the SN position \citep[the reported pointing error is $\sim 5.1^\circ, 1\sigma$, to which we should add the systematic error of 2-3$^\circ$,][]{Singer13}.

Fig.\,\ref{fig:snz} shows the data simply plotted without any fitting and considering the GRB time as the SN explosion time. The agreement, within the limits of our sparse monitoring, is remarkable. Assuming these events are really associated, GRB\,150827A would be a low-luminosity GRB, $E_{\rm iso} \sim 10^{49}$\,erg,  similar, in energy output, to the underluminous GRBs 980425 and 031203 \citep{Yamazaki03, Amati06, Ghisellini06},  and to the X-ray flashes 060218 and 100316D \citep{Campana06, Starling11}.

It would also be compatible with the luminosity function derived, e.g., in \citet{Pescalli15}. 

Although the connection of the {\it Fermi}-GBM event and the optical transient draws a credible scenario, we cannot rule out the possibility of a chance association. As an example, in Fig.\,\ref{fig:snz}, the bottom panel shows the light-curves of a standard type Ia SN 1999ee \citep{Stritzinger12}  or even with that of the peculiar type Ia SN 1991T \citep{Cappellaro2001} are also consistent with our data.

\item VSTJ60.54727-59.91890 was detected already during the raising phase in an uncatalogued galaxy probably of spiral morphology. Its light-curve is compatible with several different SN types at different redshift in the range $0.04 - 0.14$. The best fit is for a SN\,II at $z \sim 0.07$.

\item VSTJ61.20106-59.98816 was detected during the raising phase. The transient appears to be located in the outskirt of 
{PGC\,367032} (from HyperLEDA), a spiral galaxy with a bright core. No redshift is available.

\item VSTJ69.55973-64.47081 was detected in an uncatalogued spiral galaxy. The transient was at approximately constant magnitude ($r \sim 21.6$) for a couple of weeks after the GW\,150914 alert and then it was below our detection threshold at the end of our campaign.

\item VSTJ119.64230-66.71255 was also detected during the raising phase. It is located in the spheroidal galaxy 
{6dFJ0758321-664248} at redshift $z \sim 0.047$ \citep{Jones09}. The light-curve is consistent with both a SN\,Ia or a Ib/c.

\end{itemize}

\begin{figure*}
\begin{center}
\includegraphics[scale=0.60]{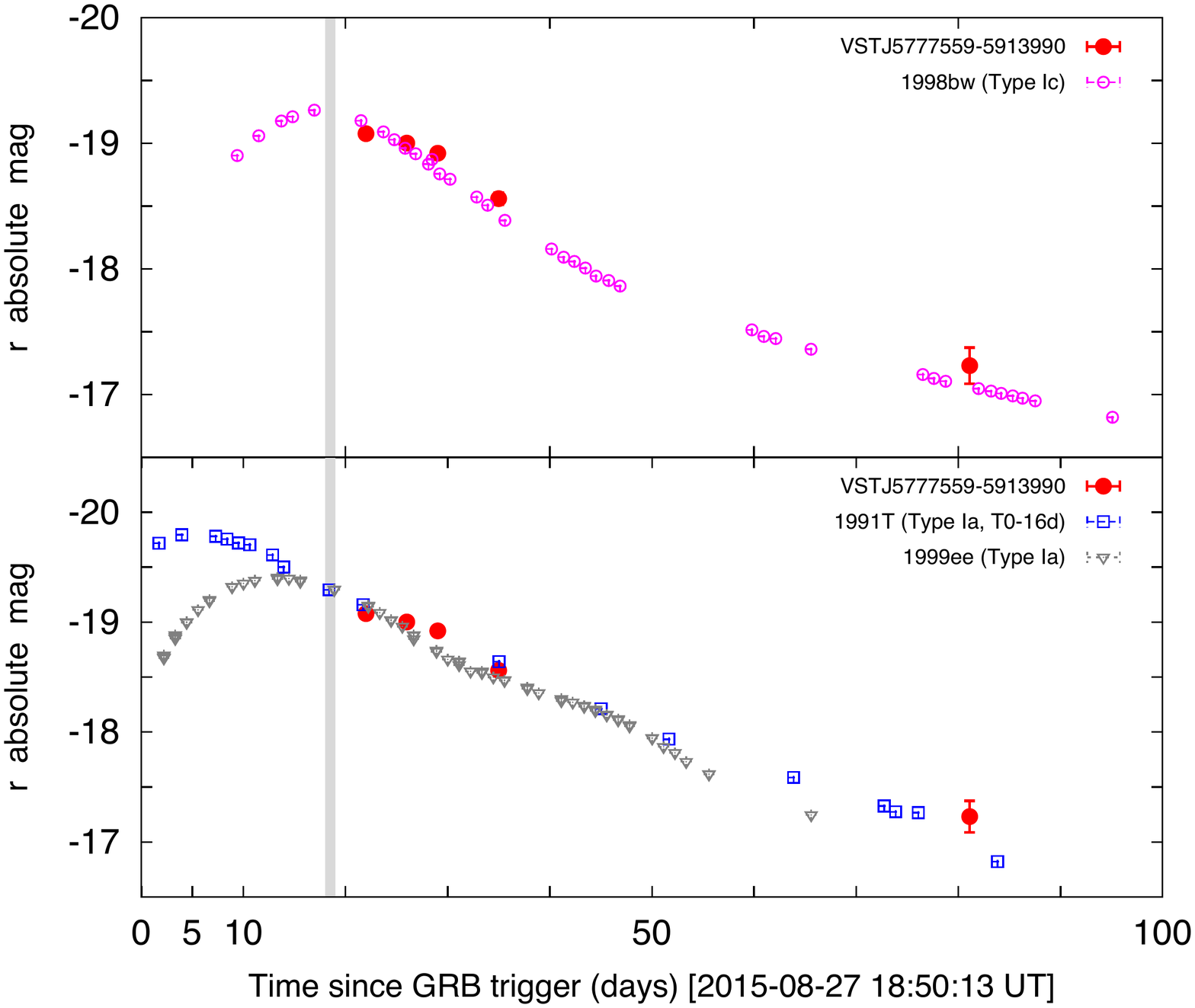}
\end{center}
\caption{{\it Top:} The light-curve of the SN candidate VSTJ5777559-5913990 
and superposed the light-curve of the hypernova prototype SN\,1998bw \citep{Iwamoto98}. The explosion time is the {\it Fermi}-GBM GRB 150827A event time, and data for the SN are simply scaled to the redshift of the likely host galaxy at $z \sim 0.11$. The agreement with the observed data is quite good. The vertical grey line identifies the GW event time. {\it Bottom:} The same data plotted with the light-curves of two SNe of the Ia family, SN\,1991T \citep{Lira98} and SN\,1999ee \citep{Stritzinger12}. The SN\,1999ee light-curve is also in reasonable agreement with the data. It is clear that without a spectroscopic confirmation, with only sparse photometric information, it is not possible to classify a SN reliably. If the {\it Fermi}-GBM event time and the optical transient are not associated even the light-curve of the peculiarly bright SN Ia as SN\,1991T can be in agreement with the observations assuming that the explosion time was about 16\,days before the (unrelated) high-energy event.  
}
\label{fig:snz}
\end{figure*}

Assuming all these objects are SNe and including the three other SNe first discovered in other surveys (we did not consider the likely AGN OGLE-2014-SN-094, Table\,\ref{tab:coords}), we count 10 SNe. This can be compared with the expected number of SNe based on the known SN rates in the local Universe, the survey area, the light curve of SNe, the time distribution of the observations, the detection efficiencies at the different epochs \citep[c.f. Sect. 5.1 of][]{Smartt16}.  For this computation we used a tool specifically developed for the planning of SN searches \citep{Cappellaro15}. We estimate an expected number of 15-25 SNe that suggest that our detection efficiency is roughly 50\%.

\subsection{GW151226}

The follow-up campaign for GW151226 was also characterized by a prompt response to the trigger and deep observations over a large sky area (see Section~\ref{obs}) Different from the follow-up campaign carried out for GW150914, the covered fields are at moderate Galactic latitude and close to the Ecliptic. In fact, the total number of analyzed sources was about an order of magnitude below the former case.

The {\tt diff-pipe} procedure
produced a list of 6310 candidates of which 3127 with high score. Performing a crosscheck of our candidate catalog with SIMBAD database gave 54 matches with known variable sources.
The candidate list shows a large number of transients that appear only at one epoch due to the large contamination from minor planets,  which was expected for the  projection of the GW151226 sky area onto the Ecliptic.
A query with Skybot\footnote{http://vo.imcce.fr/webservices/} showed a match of 3670 candidates with known minor planets within a radius of 10\arcsec.  
The {\tt ph-pipe} yielded 305 highly variable/transient
sources (after removing the known sources reported in the GAIA catalogue and the
known minor planets).
90\% of them are also part of the list provided by the {\tt diff-pipe}. Even for GW151226 most of the sources identified by the {\tt ph-pipe} and not included in the catalog produced by the {\tt diff-pipe} turned out to be real.

\subsubsection{Previously discovered Transients}

We searched in our candidate list the sources detected by the Pan-STARRS (PS) survey from Table\,1 of  \cite{Smartt16b}. Of the 56 PS objects 17 are in our survey area. Out of these, 10 ($\sim 60$\%) were identified also
by our pipelines as transient candidates. The main reason for the missing detections is the lack of proper reference images. As mentioned above, in the ESO/VST archive we could not find exposures for the surveys area of the two triggers obtained before the GW events. Therefore, we have an unavoidable bias against the detection of transients with slow luminosity evolution in the relatively short time window of our survey.
The PS candidates detected in our survey are:

\begin{itemize}
\item 
{PS16bqa} is a SN candidate first announced by \cite{Smartt16b}.

\item 
{PS15csf} was classified by the PESSTO team \citep{Harmanen15} as a SN\,II at $z \sim 0.021$.

\item 
{PS15dpn} was classified by \citet{Palazzi16} as a SN\,Ibn at $z \sim 0.1747$.

\item 
{PSN\,J02331624+1915252} was tentatively classified by \citet{Shivvers15} as a SN\,II at $z \sim 0.0135$ although the possibility it is an AGN in outburst or a tidal disruption event is not ruled out. In our images the transient was at $r \sim 20.6$.  

\item 
{PS15dom} was classified by \citet{Pan16} as a SN\,II at $z \sim 0.034$.

\item 
{PS15don} was classified by \cite{Smartt16b} as a SN\,Ia at $z \sim 0.16$.

\item 
{PS15dox} was classified by the PESSTO team \citep{Frohmaier16} as a SN\,Ia at $z \sim 0.08$.

\item 
{PS16kx} is a SN candidate proposed by \cite{Smartt16b}.

\item 
{PS15doy} was classified by \cite{Smartt16b} as a SN\,Ia at $z \sim 0.19$

\item 
{PS16ky} is a SN candidate first announced by \cite{Smartt16b}.

\end{itemize}
\begin{table*}
\begin{center}
\caption{Coordinates of the known or newly identified sources (SNe or candidate SNe) derived from the GW\,151226 follow-up campaign discussed in this section.}
\label{tab:coords2}.
\hspace*{-0.5cm}
\begin{tabular}{lllll}
\hline
\hline
\multicolumn{1}{c}{Id} & \multicolumn{1}{c}{RA} & \multicolumn{1}{c}{Dec} & Alternate Id & Note \\
     & \multicolumn{1}{c}{J2000} & \multicolumn{1}{c}{J2000} & \\
\hline
VSTJ39.73851+18.17786 & 2:38:57.24 & 18:10:40.4& PS16bqa & SN candidate\\
VSTJ36.50933+17.06122 & 2:26:02.24 & 17:03:40.4 & PS15csf & SN II, $z \sim 0.021$ \\
VSTJ38.24896+18.63528 & 2:32:59.75 & 18:38:07.0 & PS15dpn & SN Ibn, $z \sim 0.1747$ \\
VSTJ38.31767+19.25700 &	2:33:16.24 & 19:15:25.2 & PSN\,J02331624+1915252 & SN\,II?, $z \sim 0.0135$ \\
VSTJ38.69008+18.34381  & 2:34:45.62 & 18:20:37.7 & PS15dom & SN II, $z \sim 0.034$ \\
VSTJ38.84617+19.33631 & 2:35:23.08 & 19:20:10.7 & & SN candidate \\
VSTJ39.14621+18.21061 & 2:36:35.09 & 18:12:38.2 & & SN candidate \\
VSTJ39.29767+19.05561 & 2:37:11.44 & 19:03:20.2 & PS15don & SN Ia, $z \sim 0.16$ \\
VSTJ40.06271+22.53669 & 2:40:15.05 & 22:32:12.1 & PS15dox & SN Ia, $z \sim 0.08$ \\
VSTJ41.17617+22.61097 & 2:44:42.28 & 22:36:39.5 & PS16kx & SN candidate \\
VSTJ41.97567+21.77333 & 2:47:54.16 & 21:46:24.0 & PS15doy & SN Ia, $z \sim 0.19$ \\
VSTJ45.37163+28.65375 & 3:01:29.19 & 28:39:13.5 & & SN candidate \\
VSTJ46.51175+27.70492 & 3:06:02.82 & 27:42:17.7 & & SN candidate \\
VSTJ50.64421+30.60197 & 3:22:34.61 & 30:36:07.1 & PS16ky & SN candidate \\
\hline
\end{tabular}
\end{center}
\end{table*}

\begin{figure*}
\begin{center}
\includegraphics[scale=0.4]{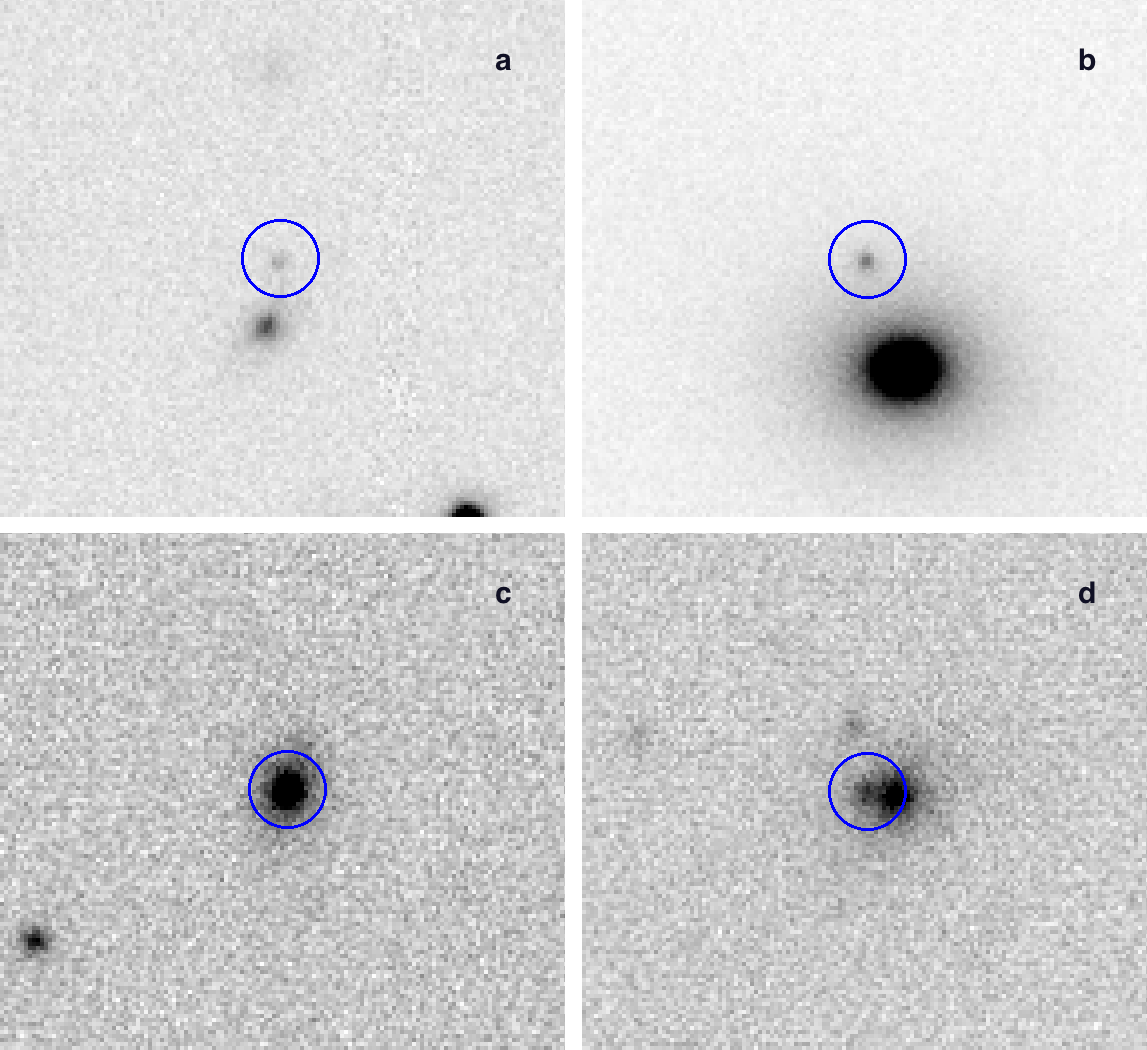}
\end{center}
\caption{A few SN candidates identified in our survey after GW151226. \textit{a}. VSTJ38.84617+19.33631 observed on 2016, Jan. 01. \textit{b}. VSTJ39.14621+18.21061 observed on 2016, Jan. 01. \textit{c}. VSTJ45.37163+28.6 observed observed on 2016, Jan. 05. \textit{d}. VSTJ46.51175+2770492 observed on 2016, Feb. 02.
In all images the showed field sizes are 30$\times$30 arcsec, North is up and East to the left.
The blue annuli represent the position identified by our pipelines.
}
\label{fig:candsunknown2}
\end{figure*}

\subsubsection{Transient candidates}

In addition, we also singled out a few objects that most likely are previously undiscovered SNe (Fig.\,\ref{fig:candsunknown2}).
\begin{itemize}
\item VSTJ38.84617+19.33631 is close to an unclassified galaxy, possibly a barred spiral seen almost edge-on. The transient was caught already in the decaying phase.

\item VSTJ39.14621+18.21061 is close to the galaxy 
{2MASXJ02363494+1812327} (from HyperLEDA) of spheroidal shape. No redshift is known and the transient was already in the decaying phase.

\item VSTJ45.37163+28.65375 is at the center of an unclassified galaxy, apparently of spheroidal shape. The transient was possibly 
identified before the maximum and showed a slow evolution during our campaign.

\item VSTJ46.51175+27.70492 is slightly off-center of the galaxy 
{2MASXJ03060262+2742176} (from HyperLEDA) of spheroidal shape. No redshift is available. The transient was brightening for the whole duration of our monitoring.

\end{itemize}

\section{Upper limits for different type of GW counterparts}
\label{upplimits}

The artificial star simulations, which use the real objects images (PSF) taken during our VST surveys and take into account the cadence of the observations, allow us to derive the detection efficiency of our search for different kind of possible optical counterparts of GW events. We derive the sensitivity distance for future VST surveys, which, in the case of non detections, can be turned into upper limits for the rate of specific kinds of events.

We took a number of proposed EM transients expected to be associated with GW sources from literature (cf. Fig.\,\ref{model-lc}). We assumed as epoch the one of the GW trigger and computed the expected light curve for each of the proposed EM counterparts. Two approaches were then followed: i)  we adopted the distance derived from the GW analysis, produced all the expected light curves at that distance and compared them with the detection upper limits at the different epochs derived  from the artificial star experiments; ii) we explored the full range of possible distances regardless of the constraint from the GW trigger. We used the detection efficiency measured by artificial star experiments to compute the probability of detection for each of the transients as a function of distance.

Figure~\ref{model-lc} shows the expected light curves assuming the distance derived from GW150914 data analysis (410 Mpc). On the same figure we show an example of our detection upper limits computed from the artificial star experiments for one of the pointings (field P31).  Only three types of transients could have been detected, namely type Ic SNe-98bw like and the long GRB viewed from an off-axis observer at all epochs, and within the first 2 epochs also a bright short GRB from a viewing angle that is equal to the jet opening angle \citep{vanEerten2011}. If we had reached a deeper threshold by one magnitude, we could have detected also the kilonova emission from a NS-NS coalescing into a hypermassive NS remnant \citep{Kasen15} during the first two epochs. All the other electromagnetic transients, at that distance, would have been far too faint to be detectable.

 \begin{figure*}
\center
\includegraphics[width=0.8\textwidth]
{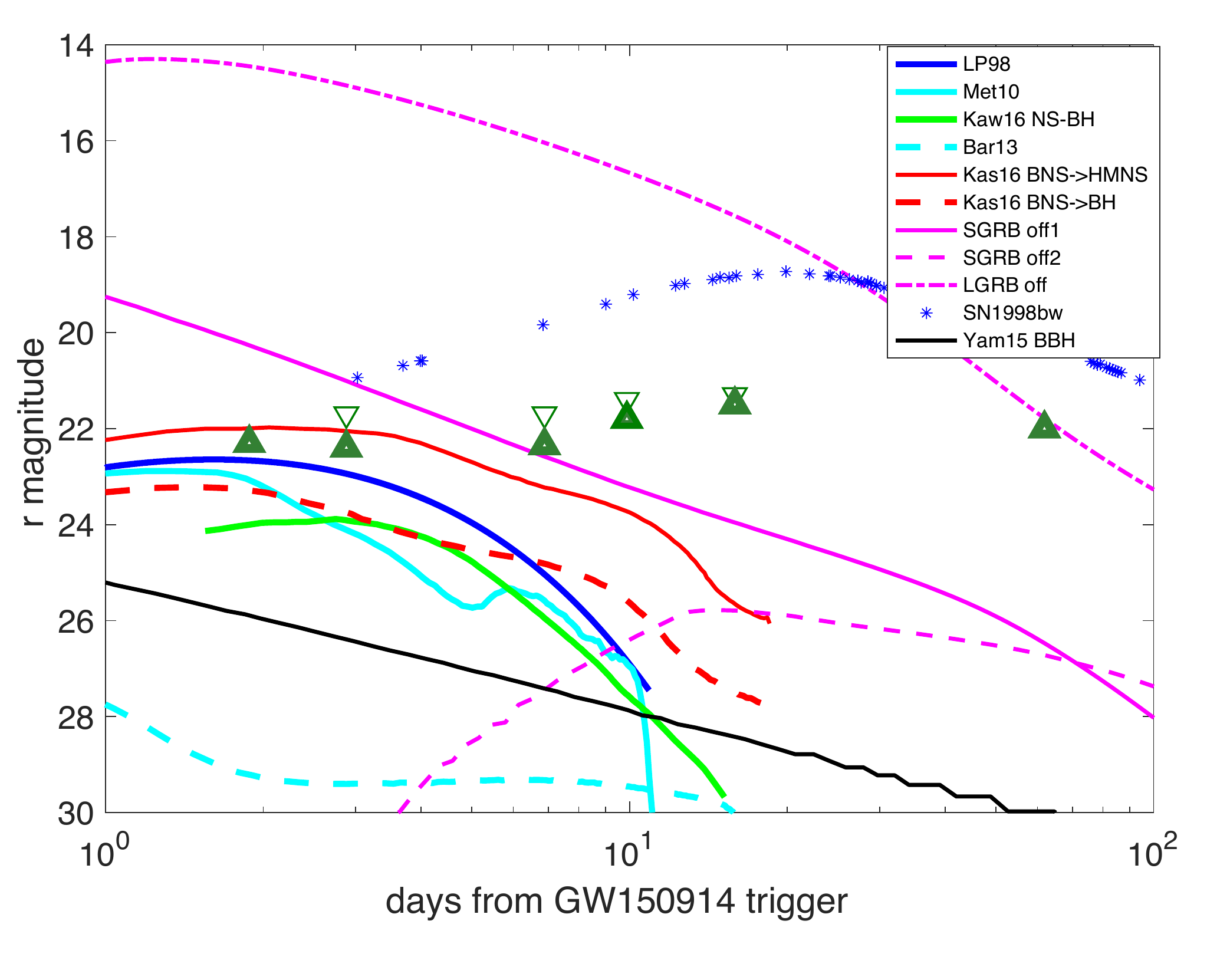}
\caption{ 
The expected fluxes (r band magnitudes) versus observed time from the GW150914 trigger, assuming several possible electromagnetic GW source emission models at the given distance of 410 Mpc, plotted against the 6 epochs VST observation 5$\sigma$ limiting magnitude ({\it dark green triangles}) and the detection upper limits
computed from artificial stars in frame P31 ({\it light green triangles}). 
{\it Blue and cyan solid line:} kilonova model from \citet{Metzger2010}, assuming a radioactive powered emission for an ejecta mass $10^{-2}$ M$_{\odot}$, outflow speed of $v=0.1c$, iron like opacities, and thermalization efficiency of 1 (cyan line) and a blackbody emission (blue line \citep{Li1998} ) with the same values of the mass and velocity.
{\it Cyan dashed line:} kilonova model from \citet{Barnes2013} assuming an ejected mass of $10^{-3}$ M$_{\odot}$ and velocity of 0.1 $c$ and lanthanides opacity. 
{\it Green solid line}: kilonova model from \citet{Kawaguchi2016} for a BH-NS merger with a BH/NS mass ratio of 3, ejected mass of 0.0256 M$_{\odot}$ and velocity $v=0.237c$, hard equation of state for the NS, and BH spin of 0.75. 
{\it Red lines}: kilonova disk-outflow models from \citet{Kasen15}, assuming accretion disc mass of 0.03 M$_{\odot}$ and a remnant hyper-massive NS (solid) or a remnant NS collapsing into a BH within 100 ms (dashed). 
{\it Purple lines}: simulated off-axis afterglow light curve \citep{vanEerten2011}, assuming a short GRB with ejecta energy of $E_{jet}=10^{50}$ erg, interstellar matter density of $n\sim10^{-3}$ cm$^{-3}$, jet half-opening angle of $\theta_{jet}\sim0.2$ rad and an observed viewing angle of $\theta_{obs}\sim0.2$ rad (solid) and $\theta_{obs}\sim0.4$ rad (dashed) and a long GRB with ejecta energy of $2\times10^{51}$ erg, $\theta_{jet}\sim0.2$ rad and an observed viewing angle of $\theta_{obs}\sim0.3$ rad  (dot-dashed line).  
{\it Blue asterisks}: SN 1998bw associated with GRB\,980425 \citep{Clocchiatti2011}. 
{\it Black solid line} : R-band emission from a BBH merging according to the model by \citet{Yamazaki2016}. 
} 

\label{model-lc}
\end{figure*}

\begin{figure*}
\center
\includegraphics[width=0.9\textwidth,angle=0]
{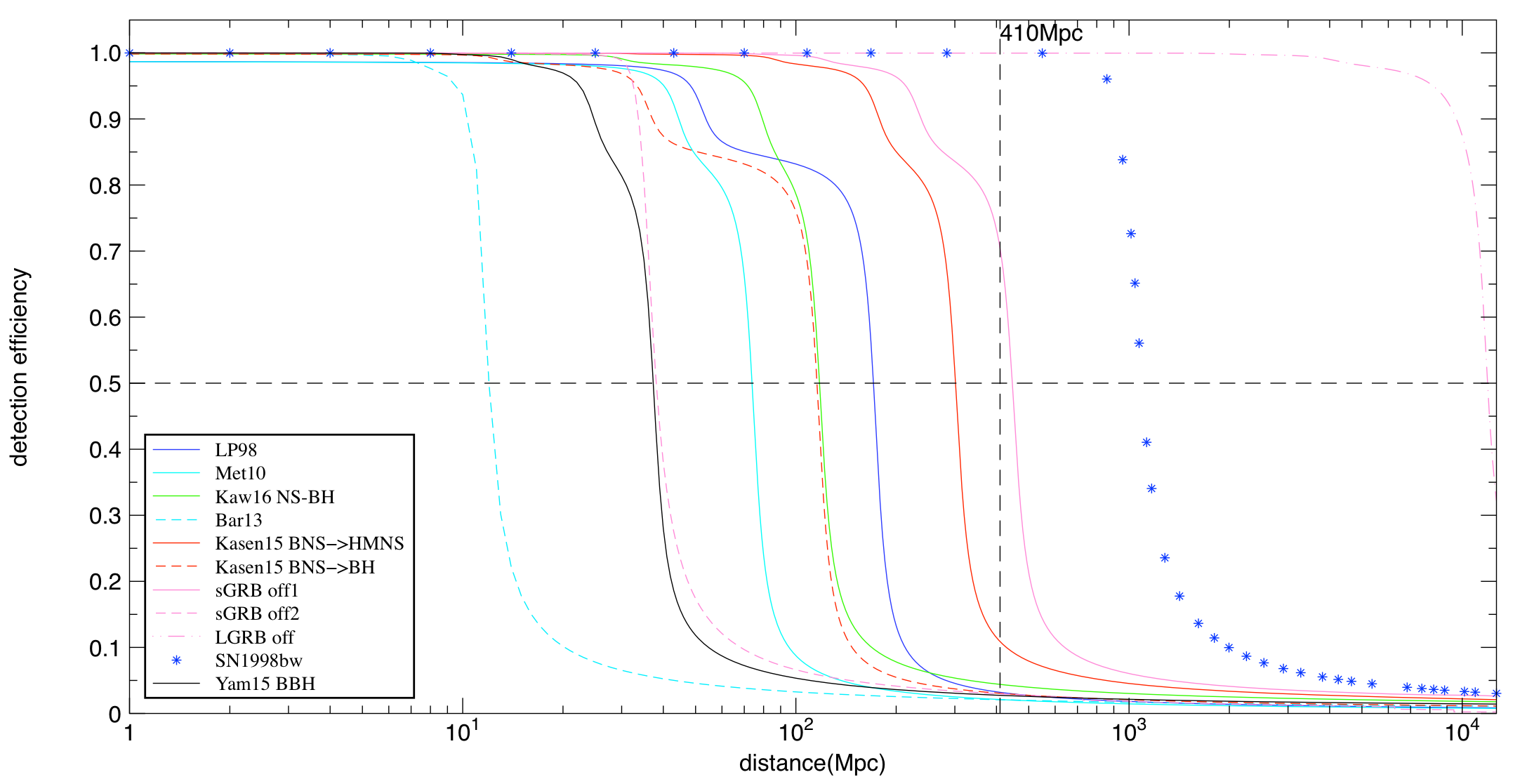}
\caption{ Detection limits for different counterpart models obtained by the artificial star experiments for the pointing P31 of GW150914. The models are described and shown as in Figure \ref{model-lc}.  The P31 field is representative of both the surveys of GW150914 and GW151226 and the curves in the figure can be considered as representative of the typical detection limit reached in the region of the sky observed for both the gravitational triggers.}
\label{det-lim}
\end{figure*}

Figure~\ref{det-lim} shows the detection efficiency as function of distance for all the models considered in figure~\ref{model-lc} and using the P31 observations of GW150914 as representative of the average properties of the VST surveys of both the GW events.
The majority  of the models associated with the merger of binary systems containing a NS (kilonova models and bright short GRBs slightly off-axis) can be detected with a detection efficiency larger than $50\%$ up to 100 Mpc. 
The expected detection rates of off-axis short GRBs in associations with GW events seems also to be promising \citep{Ghirlanda16}.
SNe and long GRBs can be detected up to distances many times larger than the detectability range of a few tens of Mpc for core collapse of massive stars by the LIGO and Virgo network. We conclude that our search for optical counterparts of GW events goes in a promising direction for securing timely observations of light curves of the expected transients within distances of the order of $\sim 100$ Mpc.

\section{Discussion and Conclusions.}
\label{concl}

GRAWITA participated in the search of the optical counterparts of the first direct detections of GWs, GW150914 and GW151226, exploiting the capabilities of the VLT survey telescope. None of the transients identified by our team can be related to the gravitational events. Nevertheless, this work made possible to verify the capabilities, reliability and the effectiveness of our project:
\begin{itemize}
\item prompt response: we started the VST observations within 23 hours after the alert for GW150914 \citep{GCN18330}, and 9 hours after the alert for GW151226 \citep{GCN18728}.

\item Observational strategy: for GW150914, VST covered  $\simeq 90$  square degrees of the GW probability sky map in the $r$ band for 6 epochs distributed over a period of 50 days. The contained probability resulted to be one of the largest obtained by optical ground based telescopes reacting to the GW150914 alert \citep{Abbott2016f}.
Concerning GW151226, the GW sky maps favoured the observation sites located in the northern hemisphere, however we were able to monitor 2 probability regions (North and South) for a total area of $\simeq 72$ square degrees for a period of 40 days.
For both the alerts, a limiting magnitude of the order of $r$ $\simeq$ $21$  mag was reached in most of the epochs.

\item Data analysis: on the basis of previous experiences in the search of GRBs and SNe, two independent pipelines have been developed. One based on source extraction and magnitude comparison between different epochs and the second on transient identification obtained through image subtraction techniques.
The two pipelines are effective and complementary. They are deeply tested and reliable, ready to be used in case of a new GW detection follow-up observational campaign during the Advanced LIGO and VIRGO network O2 run.
 
\item Transient identification:  a number of astrophysical transients has been observed and none of them can be related with plausibly reasons to the gravitational event GW150914 and GW151226.

\item By-product science: the performed survey showed the serendipitous discovery of interesting objects in the realm of the Time Domain Astronomy: peculiar supernovae and afterglows of poorly localized GRBs. For example, we suggest the connection of the supernova VSTJ57.77559-59.13990 with the {\it Fermi}-GBM  GRB\,150827A.
Further steps toward a rapid detection and characterization are critical points which, in this case, would have led to catch,
for the first time, the detection of a hypernova independently of its associated long GRB trigger.
\end{itemize}

The search for EM counterparts is very challenging due to the large sky localization uncertainties of GW signals and the large uncertainties on EM emission that GW sources may produce. The improvement of sensitivity and sky localization expected for the upcoming years, when Virgo and possibly other interferometers will join the network, will increase the chances to observe and better localize the coalescence of binary systems containing a NS, events with a significant EM signature
\citep[e.g.][]{Metzger2010,Barnes2013,Berger2014}. 

The large number of GW events expected from future runs \citep{Abbott2016b,Abbott2016d} will require an enormous EM observational effort. In the case the optimistic rates (available in recent literature) will be confirmed, the present availability of telescopes time involved in this research would not be enough to properly perform the follow-up of all the GW detections. LSST\footnote{https://www.lsstcorporation.org/science-collaborations} may partially solve the problem. The spectroscopic characterization of many candidate counterparts remains the critical bottleneck, which may be somehow mitigated by the availability of observational facilities similar to SOXS, a fast spectrograph that will be mounted at ESO-NTT \citep{Schipani2016}.

The sky areas observed for GW150914 and GW151226 reflect rather extreme properties for transients search. The GW150914 area includes the outskirt of the LMC with thousands of variable stars. The GW151226 area covers regions at low Ecliptic contaminated  by thousands of minor planets. Artificial star experiments on these fields demonstrated that the VST survey will be very valuable for hunting of the first optical counterpart, ensuring the detections of the majority of EM emission models predicted for the GW sources up to 100 Mpc.

\section*{Acknowledgements}
This paper is based on observations made with the ESO/VST. We acknowledge the usage of the VST Italian GTO time. We also acknowledge INAF financial support of the project "Gravitational Wave Astronomy with the first detections of adLIGO and adVIRGO experiments".
MB, GG, GS, MM and MS acknowledge financial support from the Italian Ministry of Education, University and Research (MIUR) through grant FIRB 2012 RBFR12PM1F.
MM and MS acknowledge support from INAF through grant PRIN-2014-14, and from the MERAC Foundation.


{\it Facility:} {VST ESO programs 095.D-0195, 095.D-0079 and 096.D-0110, 096.D-0141}. 




\bibliography{grawita}

\end{document}